\documentclass[12pt,oneside, a4paper]{revtex4}
\pdfoutput=1

\usepackage{graphicx}
\usepackage{amssymb}
\usepackage{amsmath,amssymb}
\usepackage{color}
\usepackage{setspace}
\usepackage{fancyhdr}
\usepackage{multirow,natbib}
\usepackage{caption}
\usepackage{subcaption}
\definecolor{darkblue}{rgb}{0,0,0.5}
\usepackage[colorlinks,linkcolor=blue,citecolor=blue,urlcolor=blue]{hyperref}

\allowdisplaybreaks

\usepackage[utf8]{inputenc}
\usepackage{newunicodechar}

\makeatletter
\def\UTFviii@defined#1{
	\ifx#1\relax
	!!FIXME!!%
	\else
	\expandafte‌​r#1%
	\fi
}

\makeatother

\newcommand\beq{\begin{equation}}
\newcommand\eeq{\end{equation}}

\begin{document}
%
\title{ On the possibility of observing tetraquarks in the $K^{+}$~beam }
\author{A.S.~Gerasimov}
\email{Anton.Gerasimov@ihep.ru}
\affiliation{Russian National Center Kurchatov Institute -- IHEP}
\affiliation{A.I.Alikhanyan National Science Laboratory (YerPhI) Foundation}

\author{A.K.~Likhoded}
\email{Anatolii.Likhoded@ihep.ru}
\affiliation{Russian National Center Kurchatov Institute -- IHEP}

\author{V.A.~Petrov}
\email{Vladimir.Petrov@ihep.ru}
\affiliation{Russian National Center Kurchatov Institute -- IHEP}

\author{V.D.~Samoylenko}
\email{Vladimir.Samoylenko@ihep.ru}
\affiliation{Russian National Center Kurchatov Institute -- IHEP}
\begin{abstract} 
Various models of tetraquark generation in the reaction  $ K^{+}p \rightarrow T (us;\bar{s} \bar{s})X $
are considered. The predictions for corresponding inclusive spectra were evaluated at the energy 32 and 250~GeV.
\end{abstract}

\maketitle

\section{Introduction}

The last decade has been marked by an increasing number of publications devoted to both experimental observation and various theoretical aspects of states that go beyond those that have become traditional since the discovery of the elementary fermions--quarks as the basis of hadron spectroscopy in works
\cite{Petermann:1965qlk, Gell-Mann:1964ewy}.
In the works of Gell-Mann 
\cite{Gell-Mann:1964ewy}  
and Zweig \cite{Zweig:1964ruk} 
it was pointed out that, in addition to states like $ q\bar{q} $ and $ qqq $, one should also expect
states $q q\bar{q} \bar{q}$ and $ qqqq\bar{q} $, etc.

At the same time, it should be noted that the lion's share of publications is related to states containing heavy quarks, while much less attention has been paid to exotics with light quarks.
In addition, the problems of spectroscopy (the determination of masses and widths) occupied the authors much more than the mechanisms of the production of such states in high-energy collision processes.
One of the reasons for this state of affairs is, of course, the very limited possibilities to use the potential approach and QCD perturbation theory.

In this paper, as a first step (leaving spectroscopic problems aside), we describe the possible mechanisms of inclusive production of tetraquarks and estimate the corresponding cross-sections. For definiteness, we give estimates for the case of interaction of a $K^{+}$ beam with a fixed target at beam energies of $32$~GeV (available at the IHEP U-70 accelerator) and $250$~GeV.

R. L. Jaffe in his early work \cite{Jaffe:1976ig} for the first time predicted in the framework of the MIT Bag
model a possible existence of tetraquarks in the sector of light quarks. The assumption
was based on the fact that light mesons $a_0$ and $f_0$ can be considered as a diquark--antidiquark system.
But attempts to detect a tetraquark made up of light
quarks were unsuccessful. It is possible that this is due to the phenomenon of strong mixing of these states
with hadrons of the usual type, making it difficult to interpret them as tetraquarks.
	The situation changed radically in the analysis of final states in the decays of hadrons containing heavy quarks. The first exotic state was observed in the Belle experiment in 2003, where the $X(3872)$ resonance simultaneously had the properties of $P$-wave charmonium and the molecular state of $D^*\bar{D}$ ~\cite{Belle:2003nnu}.    
	The proximity of the $X(3872)$ mass to the $D^*\bar{D}$ meson threshold causes mixing of quarkonium with the $D^*\bar{D}$ molecule, which leads to decay modes exotic for quarkonium, but makes it difficult to interpret as a tetraquark.
	
Over the past two decades in the heavy quark sector, a large number of resonant states that claim to be tetraquarks have been discovered \cite{Brambilla:2022ura}.
However, even in almost obvious situations, the resonant states observed in the decays of $B$ mesons and claiming to be tetraquarks allow a different interpretation related to the rescattering of the decay products in the final state.
	One of the latest reports on the tetraquark was published by the LHCb collaboration, which discovered a narrow $X(3875)$ resonance in the $D^*D$ system near the production threshold \cite{LHCb:2021vvq}. 
The width of this resonance is less than 1 MeV, which is unusual for interacting hadrons (for example, 
the width of the $Z_c(3985) \rightarrow D_sD^*$ resonance is $\approx$ 12~MeV).
	
In our work, we evaluate the possibility of repeating the LHCb result for a light tetraquark. Instead of a tetraquark with a heavy quark, we consider the $T_{ss}$ state produced in the $K$-meson beam, in which the $c$-quark is replaced by $s$-quark. This increases the production cross section compared to the two-gluon production of four $c$-quarks, which will make it possible to expand the energy range of the initial particles and observe the tetraquark at a lower multiplicity of secondary particles.
	
The purpose of our work is to describe the differential cross section for the production of exotic particles, diquarks (virtually) and tetraquarks. To do this, we first consider the differential cross sections for the production of ordinary particles. When calculating the cross sections, we use three models: the fusion model, the recombination model, and the quark fragmentation model. They describe the inclusive spectra of leading hadrons with good accuracy, but differ greatly in fundamentally important details.
	
The article is organized as follows:
In Section \ref{sect:model} we consider the inclusive production of $K^*$- and $\phi$-mesons at different energies in the  models listed above. 
Section \ref{sect:ss} presents a possible mechanism for the formation of tetraquarks containing $s$-quark(s).
It is noted that the decay of a tetraquark into a baryon-antibaryon pair is in compliance with the conservation of the baryon number.
In Section \ref{sect:find} predictions are made about the direction of the search for light tetraquarks.
 
\section{Models for describing the differential cross section for particle production}
\label{sect:model}	
		
\subsection{Fusion model}
	
\begin{figure}[h]
		\centering
		\includegraphics[width=0.8\linewidth]{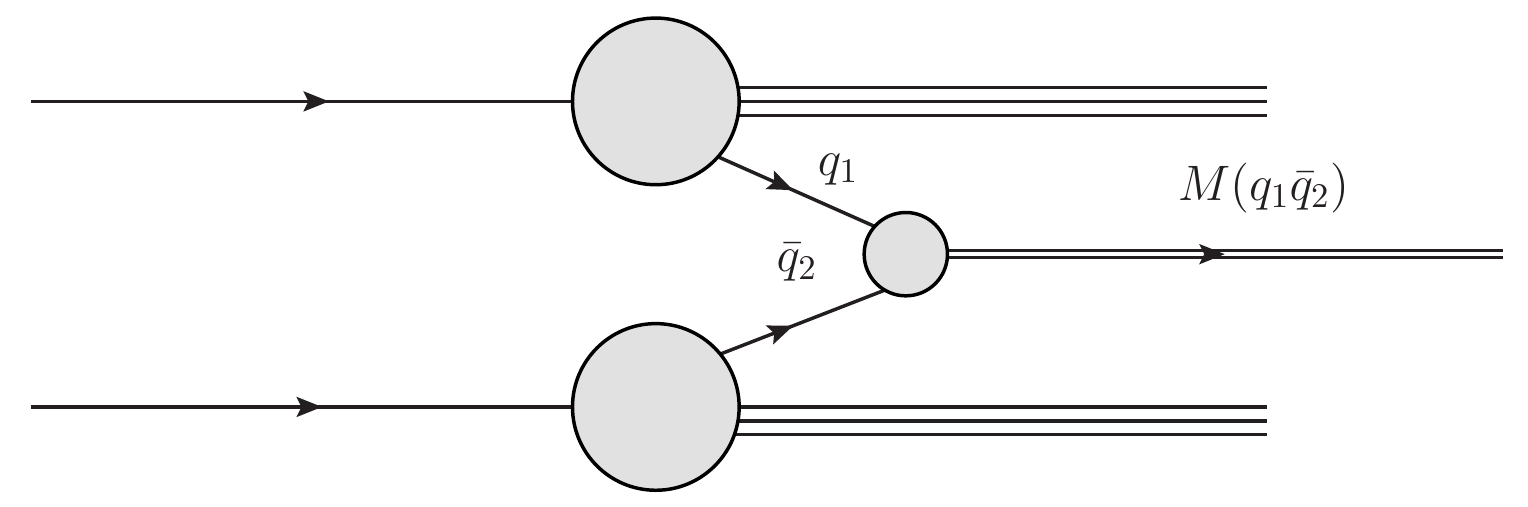}
		\caption{Diagram of the fusion process}
		\label{diag:fusion}
	\end{figure}

The fusion model is the simplest model that we used earlier \cite{Chliapnikov:1977fc} to illustrate the difference 
in the distributions of $u$ and $\bar{s}$ valence quarks in the $K^+$ meson. 
The fusion model assumes that quarks from hadrons colliding with each other are combined into a final particle.
The diagram of such a process is shown in Fig.~\ref{diag:fusion}.
The inclusive cross section for meson production in this model is as follows
	\begin{align}
		x^*\frac{d\sigma}{dx} = \frac{d\sigma}{dy} = \frac{1}{3} \frac{g^2}{4\pi} \frac{4\pi^2}{M^2} x_1 x_2 F(x_1,x_2),
	\end{align}
	where $g$ is a constant, $M$ is the mass of the final particle, $x_i$ is the momentum fraction of the initial partons , $F(x_1,x_2)$ is the sum of the products of the parton distributions of the quarks entering the final meson
	
	\begin{align}
		F = f^{A}_{q_1}(x_1)*f^{B}_{\bar{q}_2}(x_2) + f^{A}_{\bar{q}_2}(x_1)*f^{B}_{q_1}(x_2).
	\end{align}
	In the case we are considering, $A$ is a $K^+$-meson, $B$ is a proton.
Parton distributions of quarks are presented as
	
	\begin{align}
		f_{q}^{i}(x) = V_{q}^{i}(x)+S_{q}^{i}(x),
	\end{align}
	where $q$ is the quark under consideration, $i$ is the particle that contains this quark, $V$ corresponds to the valence quarks, and $S$ to the sea quark.
	In our work, as an example, we consider the $K^+ p$ interaction. The specific form of the functions $V_{q}^{i}(x)$ and $S_{q}^{i}(x)$ for quarks in the kaon and in the proton is presented in Appendix~\ref{app:part_raspr}. In doing so, we took into account the suppression coefficient $\lambda_s=0.3$ of sea strange quarks in the parton distributions of initial particles in comparison with the distributions of light quarks $u$ and $d$.
	
	The kinematics of the fusion process is as follows
	\begin{align*}
		x_1 - x_2 &= x \\
		x_1*x_2 &= \frac{M^2}{s},
	\end{align*}
where $x$ is the momentum fraction of the detected hadron.

Note that in the considered version of the fusion model, the cross section depends not only on the corresponding parton distributions, but also on the mass of the produced particle, as well as on the dimensionless constant $g^2$.
See in Appendix \ref{app:const} for specific values of the constant for vector mesons.
Considering the multiplet of vector particles $K^*$, $\rho$, $\phi$, it turns out that the constant $g^2$
for them has the same order of magnitude.
	 
\begin{figure}[h]
		\centering
		\begin{subfigure}[b]{0.475\textwidth}
			\centering
			\includegraphics[width=\textwidth]{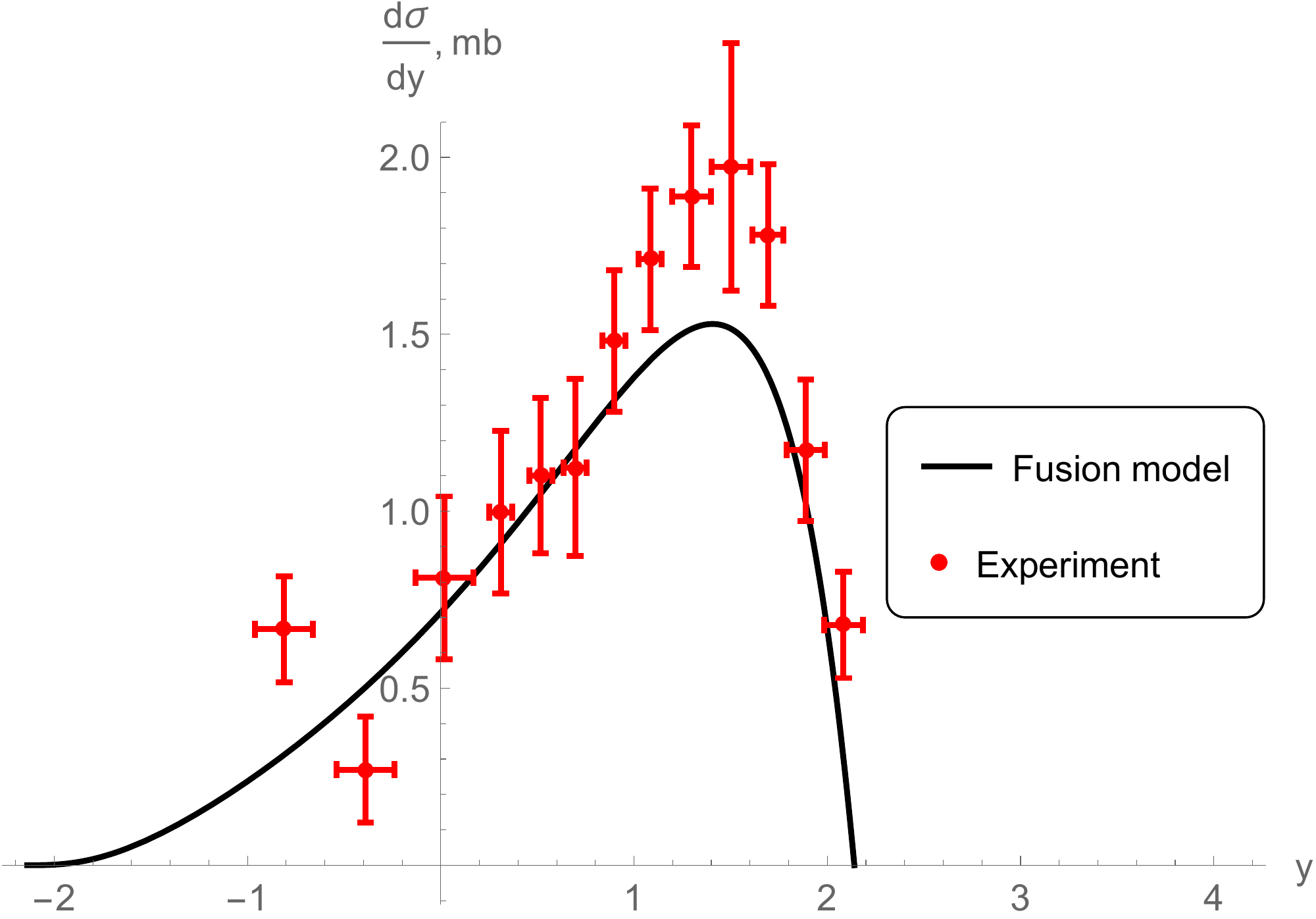}
			\caption[]%
			{{\small $K^{*0}$}}    
		\end{subfigure}
		\hfill
		\begin{subfigure}[b]{0.475\textwidth}  
			\centering 
			\includegraphics[width=\textwidth]{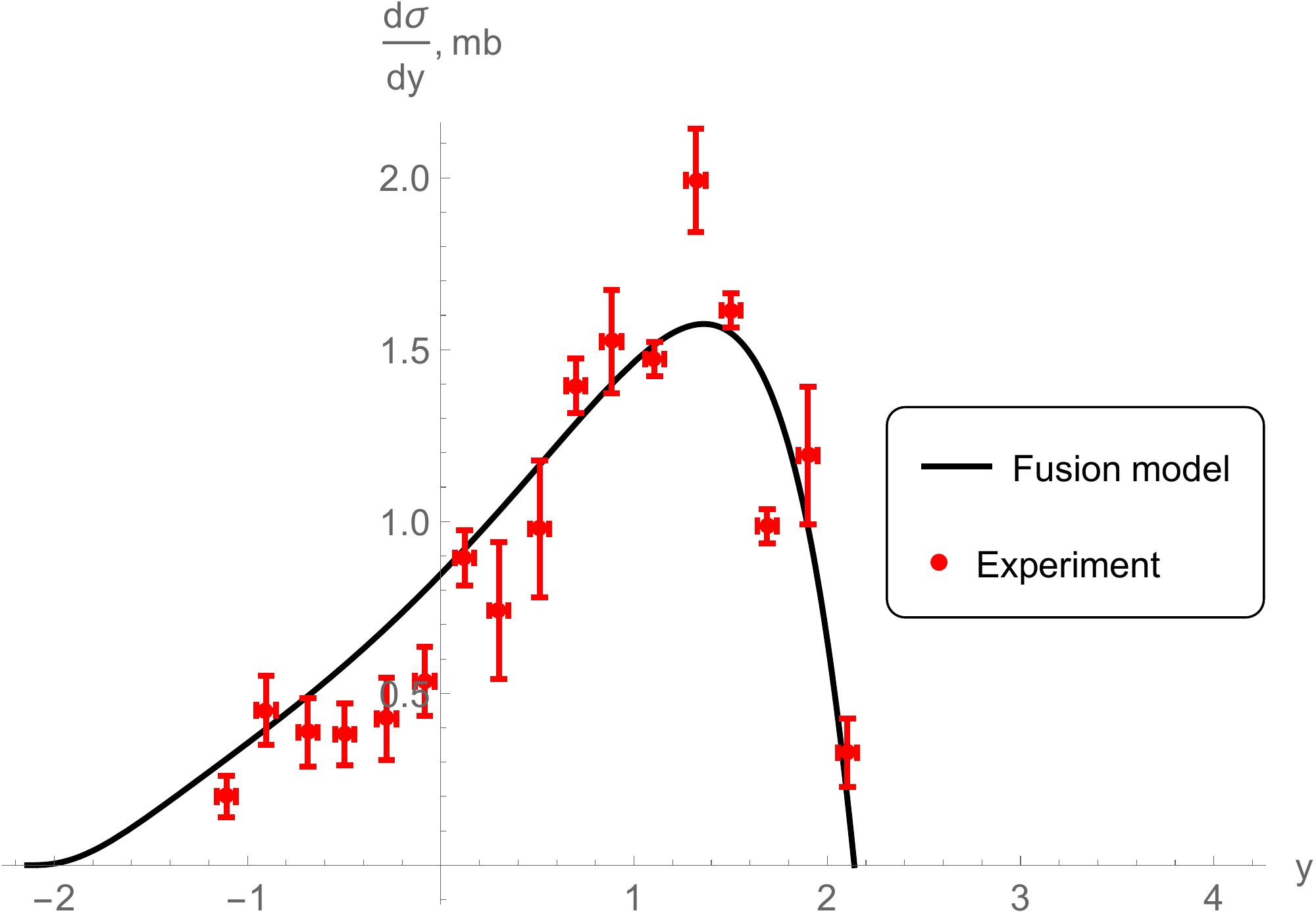}
			\caption[]%
			{{\small $K^{*+}$}}    
		\end{subfigure}
		\vskip\baselineskip
		\begin{subfigure}[b]{0.475\textwidth}   
			\centering 
			\includegraphics[width=\textwidth]{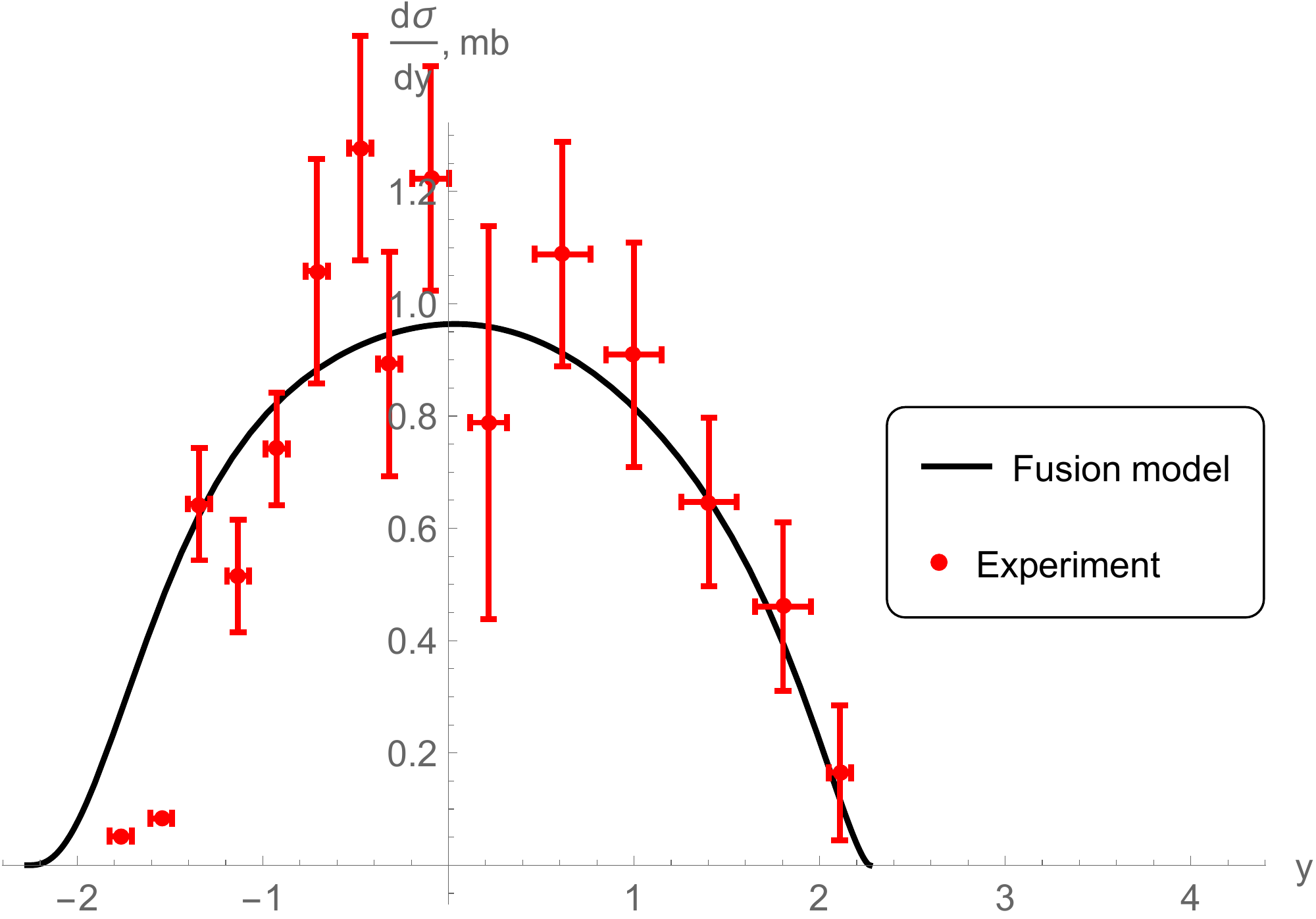}
			\caption[]%
			{{\small $\rho$}}    
		\end{subfigure}
		\hfill
		\begin{subfigure}[b]{0.475\textwidth}   
			\centering 
			\includegraphics[width=\textwidth]{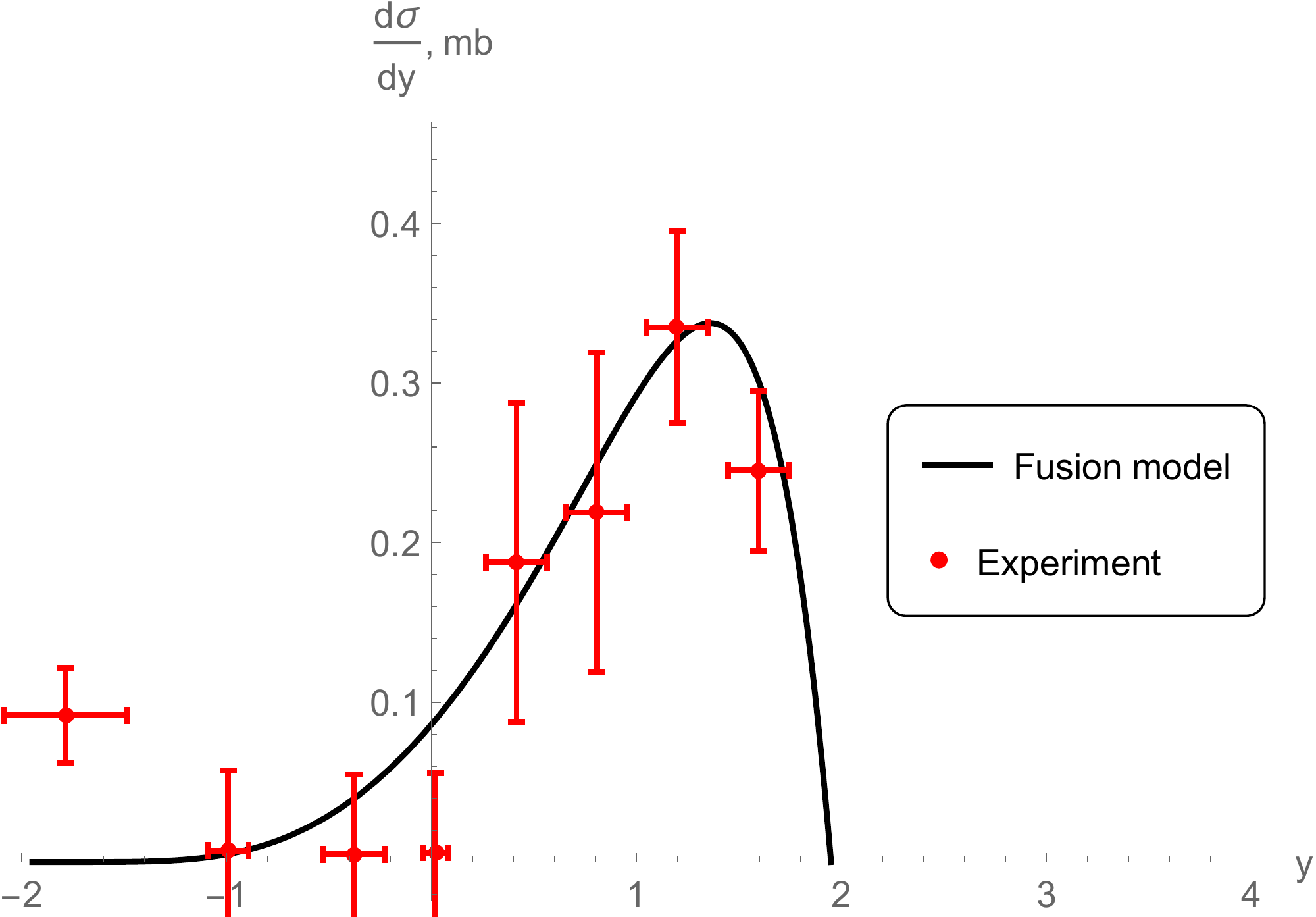}
			\caption[]%
			{{\small $\phi$}}    
		\end{subfigure}
		\caption[]
		{\small Differential cross section for the production of vector mesons.} 
		\label{fig:mesons_y}
	\end{figure}

Fig.~\ref{fig:mesons_y} shows the rapidity distributions of the cross section $y$ for vector mesons $K^{*0}$, $K^{*+}$, $\rho$,
and $\phi$ obtained in $K^+p$ collisions at $32$~GeV with the Mirabell chamber \cite{Chliapnikov:1977fc}. As expected, one can see a difference in the rapidity distributions of the $\rho$- and $\phi$-mesons, which confirms the hypothesis about the different dependence of the distribution of valence quarks in the $K$-meson. The curves presented to describe the experimental data were obtained within the framework of the quark fusion model with a fixed mass of vector mesons corresponding to the experimental values.
An important detail of the differential cross sections presented in Fig.~\ref{fig:mesons_y} is the obvious dominance of the cross sections for the production of those particles that contain valence quarks.
It can also be concluded that the valence anti-quark $\bar{s}$   in the $K^+$ meson carries away a momentum greater than the $u$ quark.
 
\begin{figure}[h]
		\centering
		\begin{subfigure}[b]{0.475\textwidth}
			\centering
			\includegraphics[width=\textwidth]{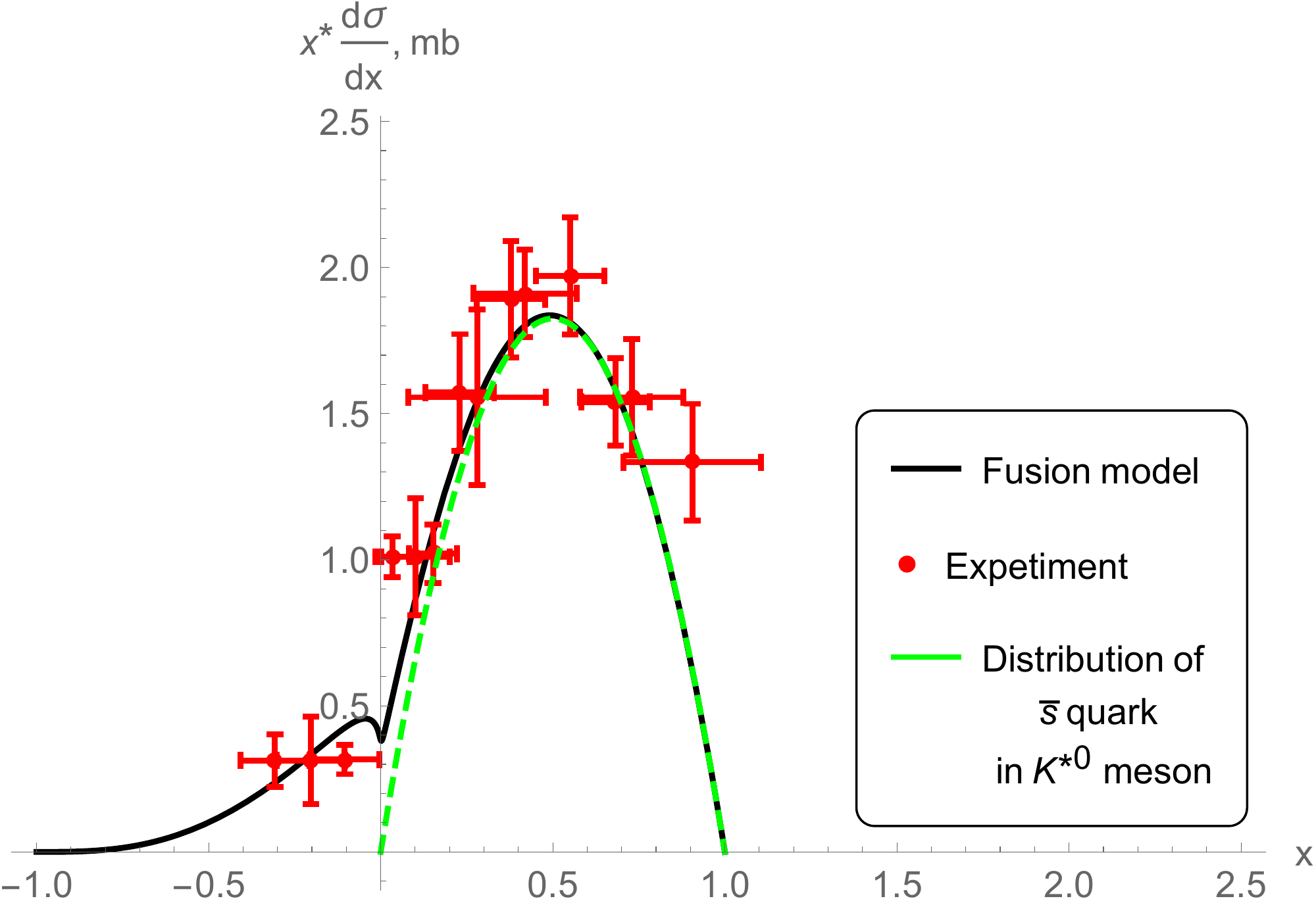}
			\caption[]%
			{{\small $K^{*0}$}}    
		\end{subfigure}
		\hfill
		\begin{subfigure}[b]{0.475\textwidth}   
			\centering 
			\includegraphics[width=\textwidth]{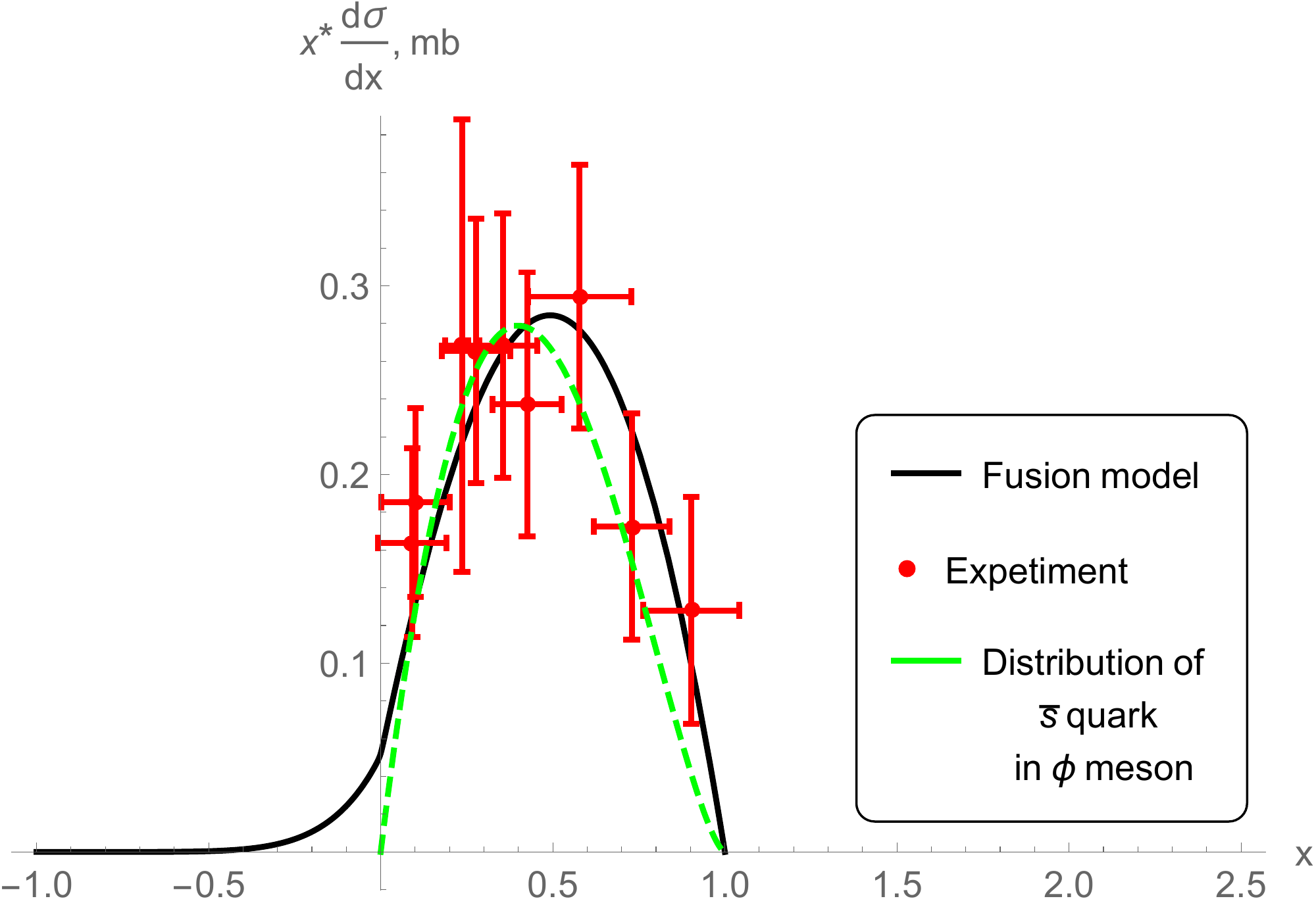}
			\caption[]%
			{{\small $\phi$}}    
		\end{subfigure}
		\caption[]
		{\small Distribution $x^{*}d\sigma / dx$  for $K^+ p$  at $ \sqrt{s}=250 $ GeV.} 
		\label{fig:mesons_xx}
	\end{figure}

\begin{figure}[h]
		\centering
		\begin{subfigure}[b]{0.475\textwidth}
			\centering
			\includegraphics[width=\textwidth]{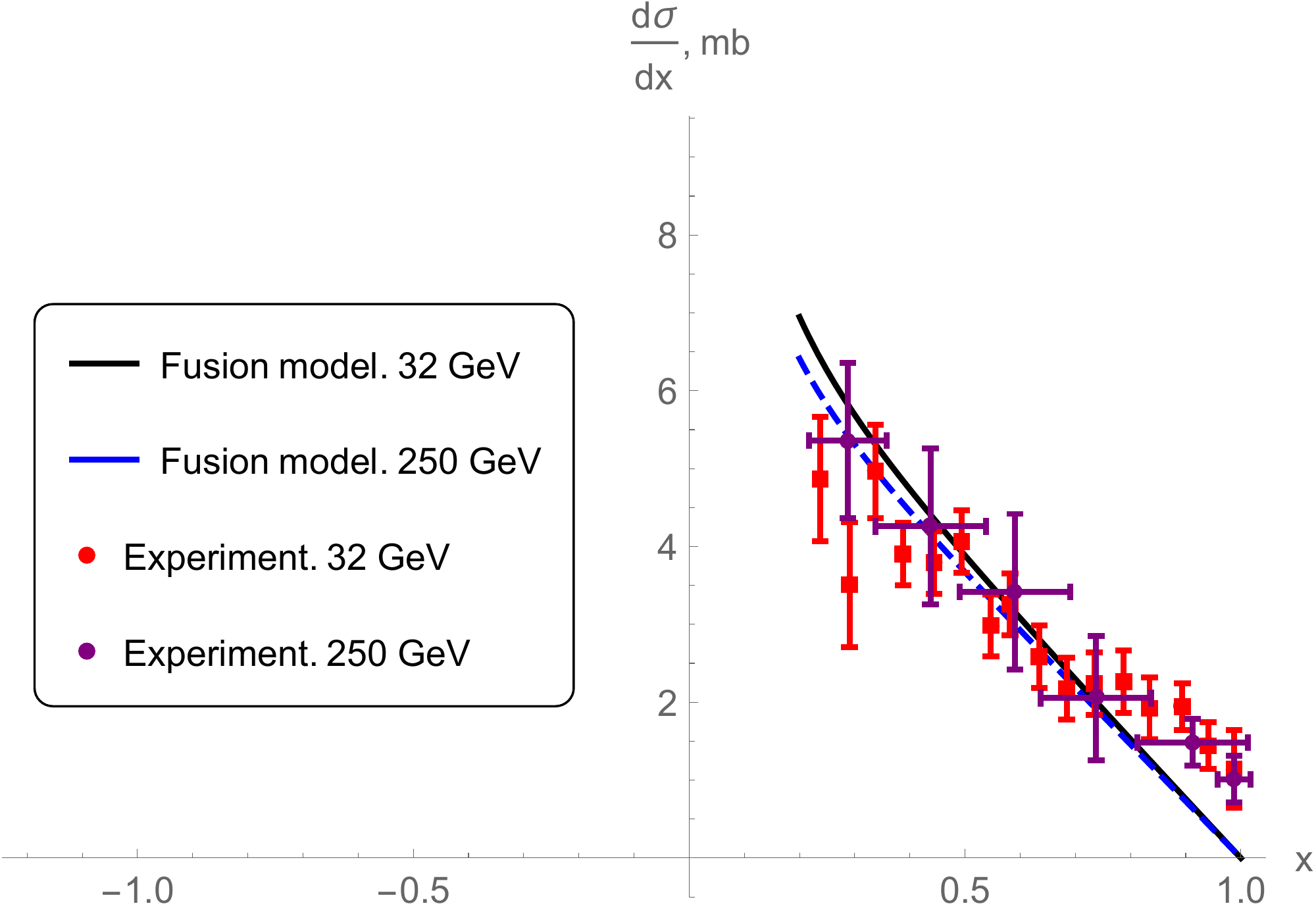}
			\caption[]%
			{{\small $K^{*0}$}}    
		\end{subfigure}
		\hfill
		\begin{subfigure}[b]{0.475\textwidth}  
			\centering 
			\includegraphics[width=\textwidth]{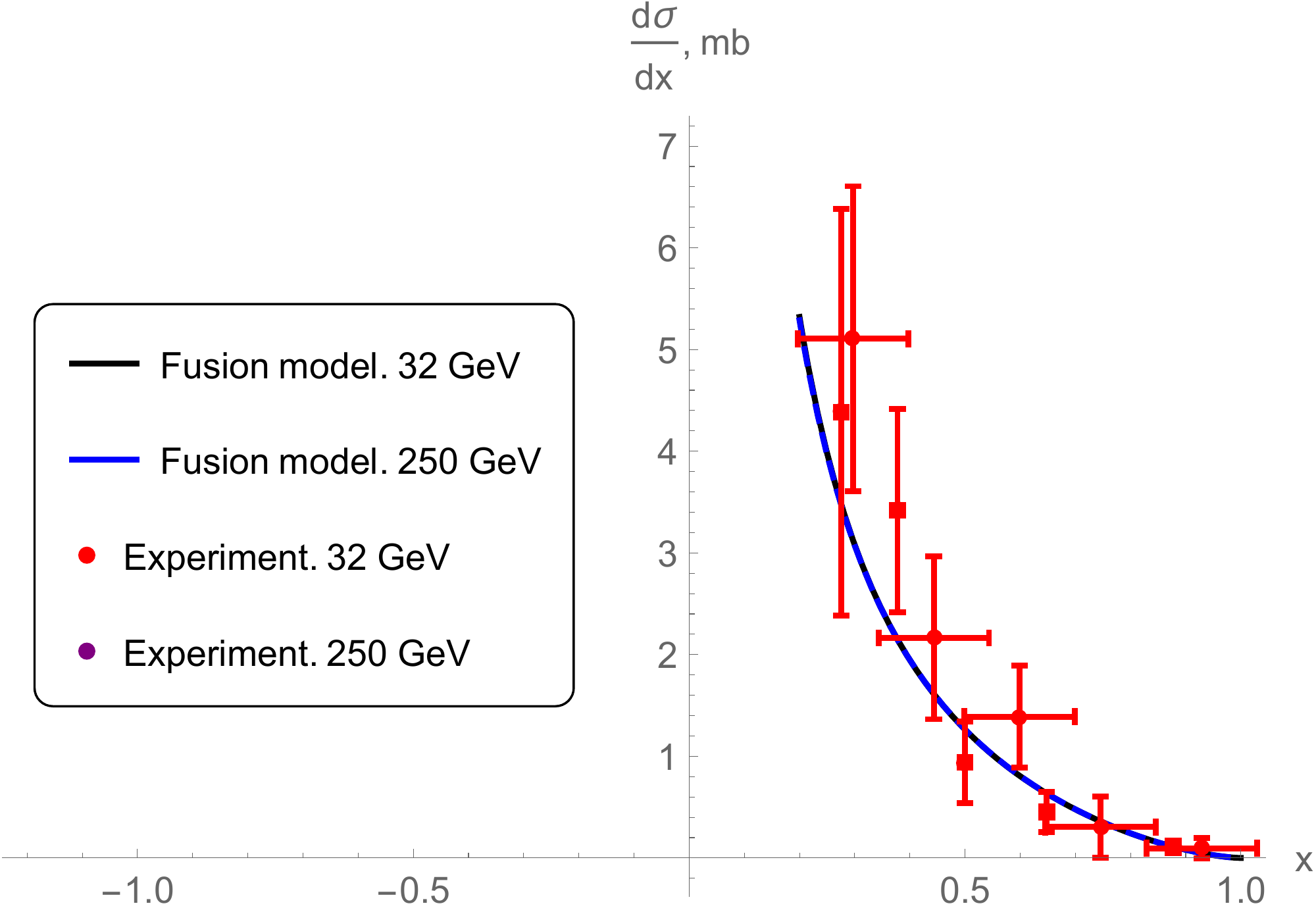}
			\caption[]%
			{{\small $\rho$}}    
		\end{subfigure}
		\vskip\baselineskip
		\begin{subfigure}[b]{0.475\textwidth}   
			\centering 
			\includegraphics[width=\textwidth]{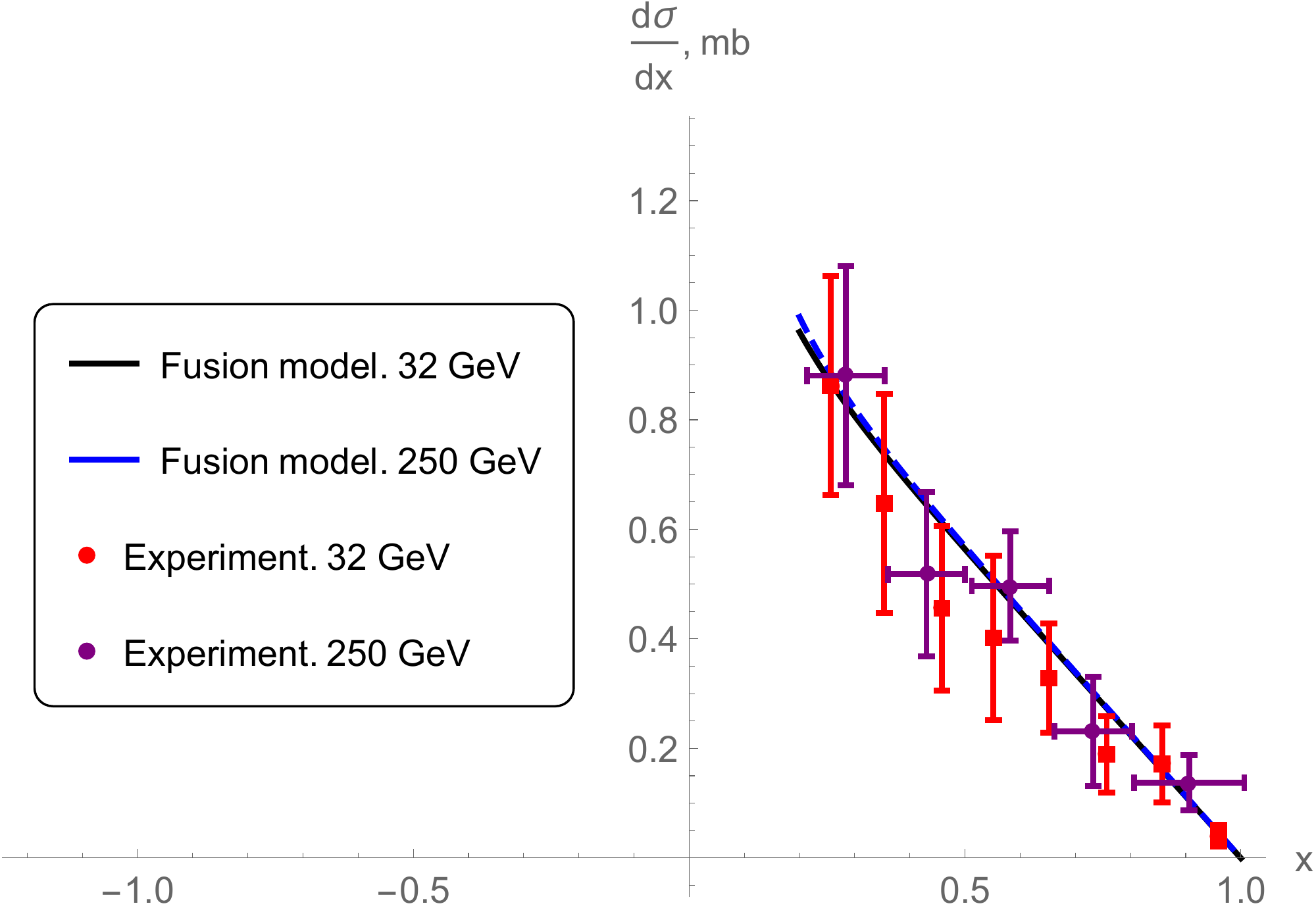}
			\caption[]%
			{{\small $\phi$}}    
		\end{subfigure}
		\caption[]
		{\small Scalling Effect.}
		\label{fig:mesons_scaling}
	\end{figure}

In addition to considering the rapidity distribution of the cross section (Fig. \ref{fig:mesons_y}), one can also consider the cross section as a function of $x$, the fraction of the momentum of the initial particle acquired by the final meson, at energies of 32 and 250~GeV. Figures~\ref{fig:mesons_xx}, \ref{fig:mesons_scaling} show such differential cross sections for meson production and their comparison with experiment \cite{EHSNA22:1989xmd}. Based on the data in Fig. \ref{fig:mesons_xx} we can conclude that the spectrum of the final particle repeats the spectrum of the valence quark.
	
	Note that for values of the variable $x > 0.2$, the spectra do not depend on the energy of colliding hadrons, i.e. scaling is observed. This phenomenon is shown in Fig.~\ref{fig:mesons_scaling}, which shows data from $K^{+}p$
at energies 32 and 250~GeV.
	
	 As it can be seen from Fig.~\ref{fig:mesons_y},\ref{fig:mesons_xx},\ref{fig:mesons_scaling}, the fusion model gives a satisfactory description
experimental data, emphasizing the violation of $SU(3)$--symmetry, which is expressed in different distributions of valence quarks in $\pi$ and $K$ mesons~\cite{Chliapnikov:1977fc}.
The disadvantage of the considered model is the neglect of the wave function of the produced particle, i.e. the constant $g^2$ is not calculated, but found from the experiment.
Next, we consider the recombination model and the fragmentation model, which take into account the wave function of the final particle.
\subsection{Recombination model}
	
The recombination model has shown itself well in describing the inclusive differential spectra of $D$ mesons produced in $\pi^-p$ collisions \cite{Likhoded:1997bm}.
Within the framework of this model, it is assumed that quarks from the initial hadron recombine into the final particle. In this paper we consider the recombination of quarks from the $K^+$ meson using its structure function. Within the framework of this model, one can also consider the process of baryon formation, but this requires taking into account the momenta of three quarks, which greatly complicates the problem.
	
	According to \cite{Likhoded:1997bm}, the differential production cross section in the recombination model is described as
	\begin{align}
		x^* \frac{d\sigma}{dx} = A^{r} \int \int x_1 x_2 \frac{d^2\sigma}{dx_1 dx_2}R(x_1, x_2, x)\frac{dx_1}{x_1}\frac{dx_2}{x_2},
		\label{eq:recomb_likhoded}
	\end{align}
	where $x_i$ is the momentum fraction of the initial partons, $x$ is the momentum fraction of the final hadron, 
	$$x_1 x_2 \frac{d^2\sigma}{dx_1dx_2}$$
	is the double differential cross section for the production of quarks in the initial hadron, $A^{r}$~is a constant that includes information about the total cross section for the production of a~particle. The recombinator $R$ has the form
	\begin{align}
		R(x_1, x_2,x) &= \frac{\Gamma(2-\beta_1-\beta_2)}{\Gamma(1-\beta_1)\Gamma(1-\beta_2)}\rho(\xi_1, \xi_2)\delta(1-\xi_1-\xi_2), \\
		\rho(\xi_1,\xi_2) &= \xi_1^{1- \beta_1}\xi_2^{1-\beta_2}, \nonumber
		\label{eq:recombinator}
	\end{align}
where $\xi_i = \frac{x_i}{x}$, $\beta_i$ is the Regge trajectory intercept  for two quarks in the final meson.
The physical meaning of the $R$ function is the wave function of the produced hadron in the infinite-momentum system \cite{Kartvelishvili:1985ac}
 	
The double differential cross section for quark production in \cite{Likhoded:1997bm} is taken from experiment. We also assume that
	\begin{align}
		\frac{d^2\sigma}{dx_1dx_2} = \sigma_0 f_{q \bar{q}}(x_1, x_2),
	\end{align}
where $\sigma_0$ is the cross section for the production of a pair of quarks in the $K^+$ beam, and $f_{q \bar{q}}(x_1, x_2)$ is the two-particle distribution of quarks in the initial particle \cite{ Likhoded:1997bm}. We consider the recombination of the $K^+$-meson in $\phi$- and
$K^{*0}$ mesons. In this case, the valence and sea quarks recombine in the $K^+$ meson, and
	\begin{align}
		f_{q\bar{q}}(x_1, x_2)=f_{VS}(x_1, x_2) = \frac{\Gamma(2+\gamma_0-\alpha_1-\alpha_2)}{\Gamma(1-\alpha_1)\Gamma(1-\alpha_2)\Gamma(\gamma_0)} x_1^{-\alpha_1}x_2^{-1}(1-x_1-x_2)^{n_v}(1-x_2)^{k}.
		\label{eq:dvuhraspr}
	\end{align}
	Then (\ref{eq:recomb_likhoded}) assumes the form
	\begin{align}
		x^* \frac{d\sigma}{dx} = A^{r} \sigma_0 \int \int dx_1dx_2 f_{q\bar{q}}(x_1, x_2)R(x_1, x_2, x),
		\label{eq:ourrecomb}
	\end{align}
where $\alpha_1$ is the Regge trajectory intercept at zero for the valence quark, $n_v$, $k$ are parameters determined according to \cite{Likhoded:1997bm}. For the case of $\phi$-meson production $n_v = 1.5$, $k = 3.5$, 
for the case of $K^{*0}$-meson production $n_v = 1 $, $k = 3.5$

\begin{figure}[h]
		\centering
		\begin{subfigure}[b]{0.475\textwidth}
			\centering
			\includegraphics[width=\textwidth]{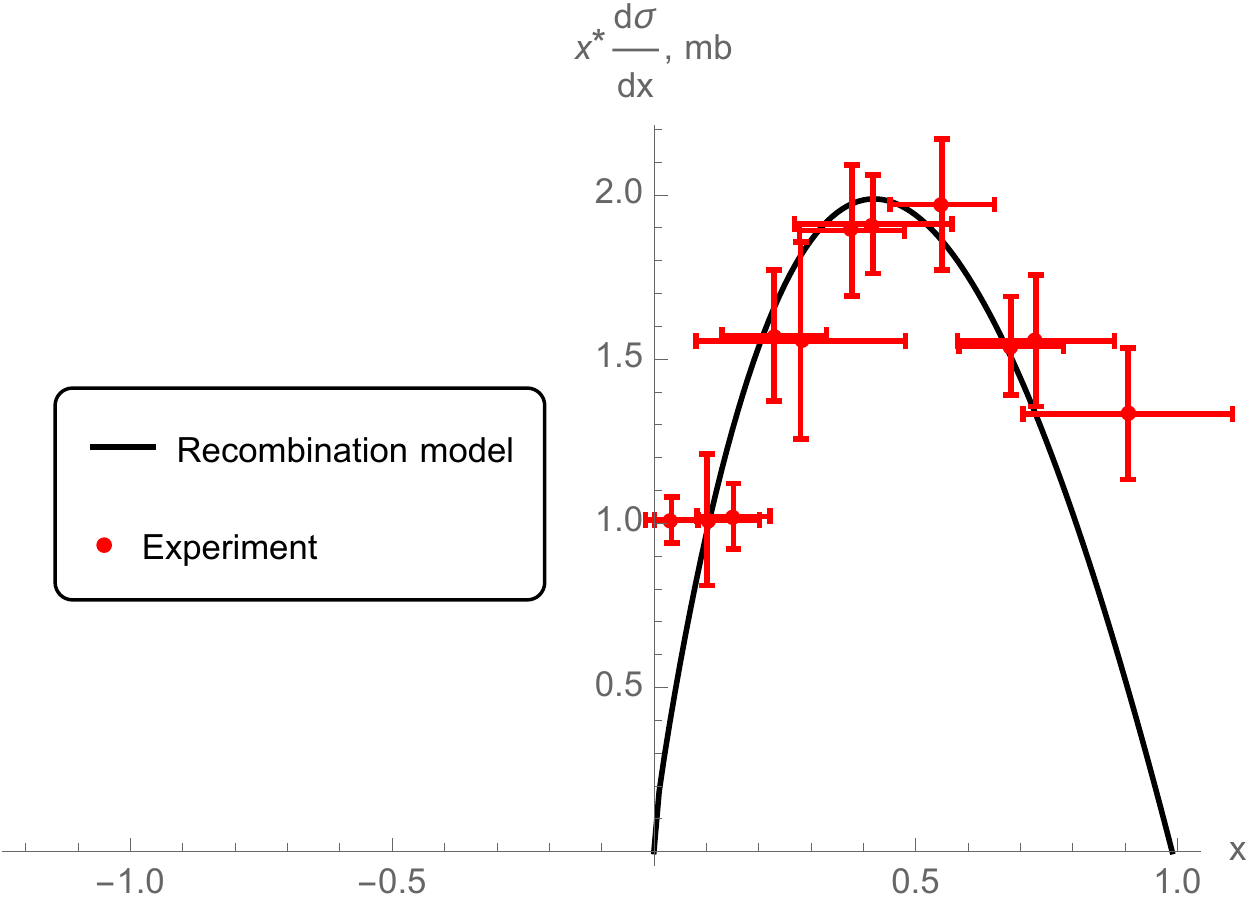}
			\caption[]%
			{{\small $K^{*0}$}}    
		\end{subfigure}
		\hfill
		\begin{subfigure}[b]{0.475\textwidth}  
			\centering 
			\includegraphics[width=\textwidth]{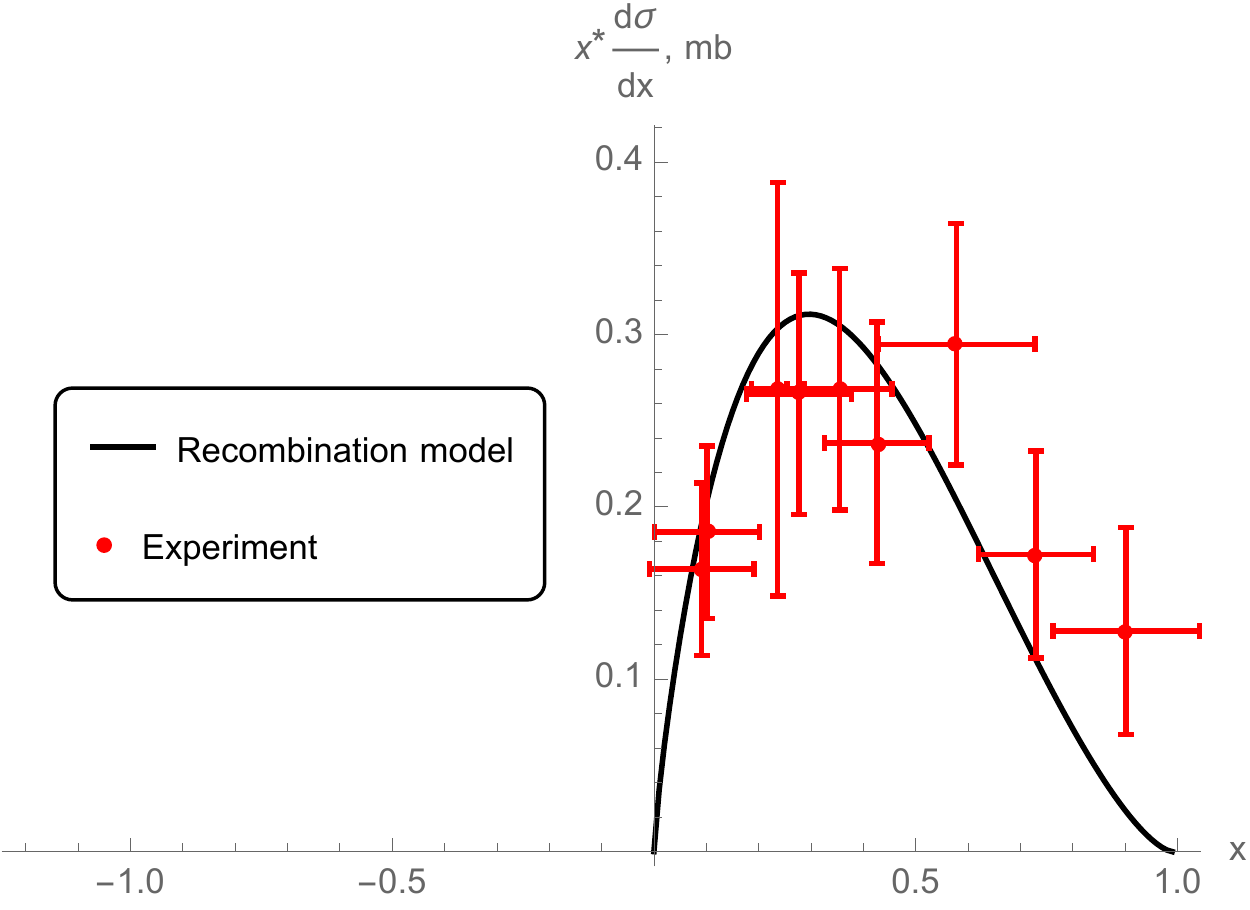}
			\caption[]%
			{{\small $\phi$}}    
		\end{subfigure}
		\caption[]
		{\small Differential cross section of $K^{*0}$ mesons (a) and $\phi$ mesons (b) at an energy of 250~GeV
(experimental data shown in red).
The description in the recombination model is shown in black.} 
		\label{fig:recomb}
	\end{figure}

Given (\ref{eq:recombinator}) and (\ref{eq:dvuhraspr}), the formula (\ref{eq:ourrecomb}) becomes
	\begin{align}
		x^* \frac{d\sigma}{dx} = B  \int \int dx_1 dx_2 x_1^{-\alpha_1}x_2^{-1}(1-x_1-x_2)^{n_v}(1-x_2)^{k} \xi_1^{1-\beta_1}\xi_2^{1-\beta_2}\delta(1-\xi_1-\xi_2),
	\end{align}
	where
	\begin{align*}
		B=A^{r}\sigma_0 \frac{\Gamma(2-\beta_1-\beta_2)}{\Gamma(1-\beta_1)\Gamma(1-\beta_2)}\frac{\Gamma(2+\gamma_0-\alpha_1-\alpha_2)}{\Gamma(1-\alpha_1)\Gamma(1-\alpha_2)\Gamma(\gamma_0)}. 
	\end{align*}
	Substituting expressions for $\xi_i$ into (\ref{eq:ourrecomb}) we get
	\begin{align}
		x^* \frac{d\sigma}{dx} = B  \int \int dx_1 dx_2 x_1^{-\alpha_1}x_2^{-1}(1-x_1-x_2)^{n_v}(1-x_2)^{k} \big( \frac{x_1}{x}\big)^{1-\beta_1} \big( \frac{x_2}{x}\big) ^{1-\beta_2}\delta(1-\xi_1-\xi_2).
	\end{align}
It is convenient to rewrite this expression as
	\begin{align}
		x^* \frac{d\sigma}{dx} = B  \int \int dx_1 dx_2 (1-x_1-x_2)^{n_v}(1-x_2)^{k} x_1^{1-\beta_1-\alpha_1} x_2^{-\beta_2}x^{\beta_1+\beta_2-2} \delta(1-\xi_1-\xi_2)
	\end{align}
	After integrating the $\delta$-function, we have
	\begin{align}
		x^* \frac{d\sigma}{dx} = B  \int_0^x  dx_1 \frac{(1-x)^{n_v}(1-x+x_1)^{k} x_1^{1-\beta_1-\alpha_1} (x-x_1)^{-\beta_2}}{x^{1-\beta_1-\beta_2}}.
		\label{eq:recombfinal}
	\end{align}
	Variants $K^+ \to \phi$ and $K^+ \to K^{*0}$ differ only in the set of constants:
	 $$\alpha_1=0, n_v=1.5, k=3.5, \beta_1=0, \beta_2=0$$
	 $$\alpha_1=0, n_v=1, k=3.5, \beta_1=0, \beta_2=1/2.$$ 
	 The equation~(\ref{eq:recombfinal}) becomes for the first case as
	\begin{align}
		x^*\frac{d\sigma}{dx}=\int_0^x dx_1 \frac{(1-x)^{1.5}*x_1*(1-x+x_1)^{3.5}}{x},
	\end{align}
%
and for the second case as
	\begin{align}
		x^*\frac{d\sigma}{dx}=\int_0^x dx_1 \frac{(1-x)*x_1*(1-x+x_1)^{3.5}}{\sqrt{x}\sqrt{x-x_1}}.
	\end{align}
In Fig.~\ref{fig:recomb} presents the differential cross sections for the production of $K^{*0}$ and $\phi$ in the $K^+$-beam.
As we can see from this figure, the recombination model describes the experiment well at small $x$, but in the region of $x>0.6$ it gives predictions that are systematically softer than the real spectrum.

Note that this model depends on the value of the constant $\gamma_0$ associated with the contribution of sea quarks, and which we consider to be the same for all mesons. If we take into account that a bound state of two quarks is produced, which does not have time (or does not have enough time) to acquire a gluon "coat" , then the value of $\gamma_0$ should be respectively taken equal to zero or reduced, which will make the spectrum at large $x$ more close to the experimental data.
	
As it can be seen, the recombination model makes it possible to describe the production of mesons, but gives a softer, in comparison with the experiment,
spectrum for large values of $x$. The essential drawbacks of the model are the lack of dependence on energy and the uncertainty of the contribution of the recombination mechanism to particle production. In addition, if we consider the production of particles with heavy quarks $c$ and $b$ in this model, then it is necessary to take into account the cross section for the production of sea quarks, which will affect the value of the total cross section.
%
%
%
\subsection{Fragmentation model}
	
The quark fragmentation model assumes that a valence quark from an initial hadron fragments into a final hadron,
for example, a $\bar{s}$ quark from a $K^+$ meson fragments into a $\phi$ meson.
If the final hadron carries away a fraction of the momentum $x$ of the momentum of the initial quark $y$, 
then the fragmentation probability
\begin{align}
		\frac{d\sigma}{dx} (x) = A^{f}   \int_x^1 dy \  f^q(y) \phi^{q \to H}\big( \frac{x}{y} \big) , 
	\end{align}
where $A^{f}$ is a constant that includes information about the total particle production cross section, $f^{q}(y)$ is the distribution of the valence quark in the initial particle, and $\phi^{ q \to H}(x)$ is the wave function of a finite hadron related to the distributions of quarks in it.
	
The distribution of the valence quark in the initial particle, according to \cite{Kuti:1971ph}, is as follows
	\begin{align}
		f^{q} (y) = \frac{\Gamma(2+\gamma_m-\alpha_v-\alpha_2)}{\Gamma(1-\alpha_v)\Gamma(1+\gamma_m-\alpha_2)} y^{-\alpha_v} * (1-y)^{-1+\gamma_m + 1 - \alpha_2}, 
		\label{eq:frag:fqy}
	\end{align}
	where $\alpha_v$ is the Regge trajectory intercept for the valence quark , $\alpha_2$ is the Regge trajectory intercept for the second quark in the initial meson. If we restrict ourselves to the approximation of valence quarks, then in (\ref{eq:frag:fqy}) 
	the parameter $\gamma_m=0$.

The quark fragmentation function $\phi^{q \to H}$ is related to the distribution of the valence quark in a finite hadron and is already determined by the wave function of this hadron in the infinite momentum frame, the form of which \cite{Kartvelishvili:1985ac} we calculate according to the same rules \cite{Kuti:1971ph} as the structure function of the initial hadron.
	
In the case of fragmentation into a meson, after replacing $\frac{x}{y} = z$ we get
	\begin{align}
		\phi^{q \to H} (z) =\phi^{q \to M} (z)= \frac{\Gamma(2+\gamma_m-\alpha_v-\beta)}{\Gamma(1-\alpha_v) \Gamma(1+\gamma_m-\beta)}z^{-\alpha_v} * (1-z)^{-1+\gamma_m + 1-\beta}
		\label{eq:frag_mes_init},
	\end{align}
where $\beta$ is the intercept of the Regge trajectory for the second quark in the final meson.

We restrict ourselves to the approximation of valence quarks, assuming that when a meson is produced, it does not have time to acquire a gluon "coat", so in (\ref{eq:frag_mes_init}) $\gamma_m=0$ and
	\begin{align}
		\phi^{q \to M} (z)= \frac{\Gamma(2+\gamma_m-\alpha_v-\beta)}{\Gamma(1-\alpha_v) \Gamma(1+\gamma_m-\beta)} z^{-\alpha_v} * (1-z)^{-\beta}.
		\label{eq:frag_mes}
	\end{align}
	In the case of fragmentation into a baryon
	\begin{align}
		\phi^{q \to H} (z) = \phi^{q \to B} (z)= \frac{\Gamma(\gamma_b + 3 - \alpha_v - \beta_1 - \beta_2)}{\Gamma(1-\alpha_v)\Gamma(\gamma_b+2-\beta_1-\beta_2)}z^{-\alpha_v} * (1-z)^{-1+\gamma_b + 1-\beta_1 + 1 - \beta_2},
	\end{align}
	where $\beta_i$ is the Regge intercept for the two remaining quarks in the final baryon.

\begin{figure}[h]
		\centering
		\begin{subfigure}[b]{0.475\textwidth}
			\centering
			\includegraphics[width=\textwidth]{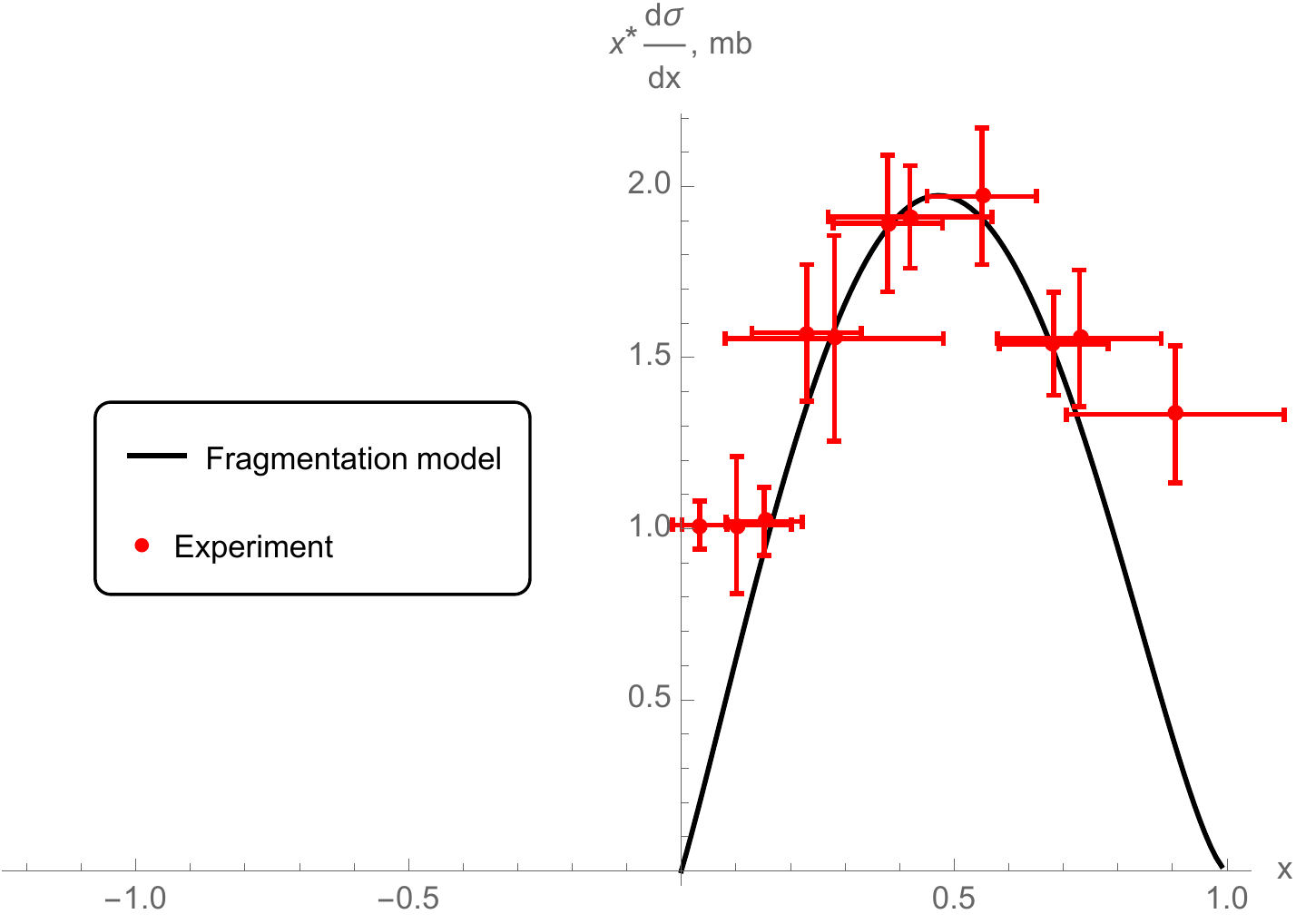}
			\caption[]%
			{{\small $K^{*0}$ 250 GeV}}    
		\end{subfigure}
		\hfill
		\begin{subfigure}[b]{0.475\textwidth}  
			\centering 
			\includegraphics[width=\textwidth]{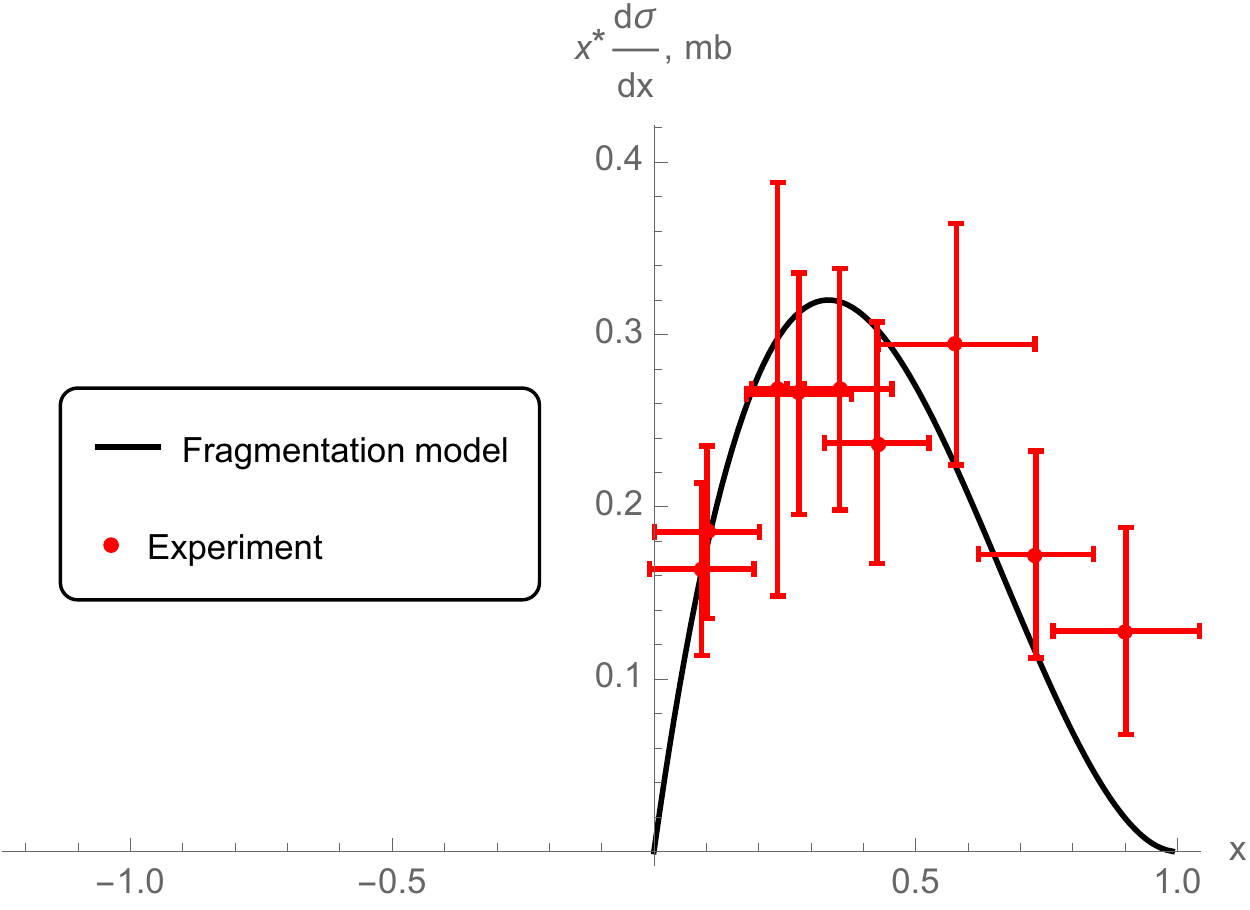}
			\caption[]%
			{{\small $\phi$ 250 GeV}}    
		\end{subfigure}
		
		\caption[]
		{\small Differential cross section of $x^*\frac{d\sigma}{dx}$ $K^{*0}$- and $\phi$-mesons,
((a) and (b) respectively). Experimental data are shown in red,
the description in the fragmentation model is shown in black.}
		\label{fig:frag_mes_xdsdx}
	\end{figure}

\begin{figure}[h]
		\centering
		\begin{subfigure}[b]{0.475\textwidth}
			\centering
		\includegraphics[width=\textwidth]{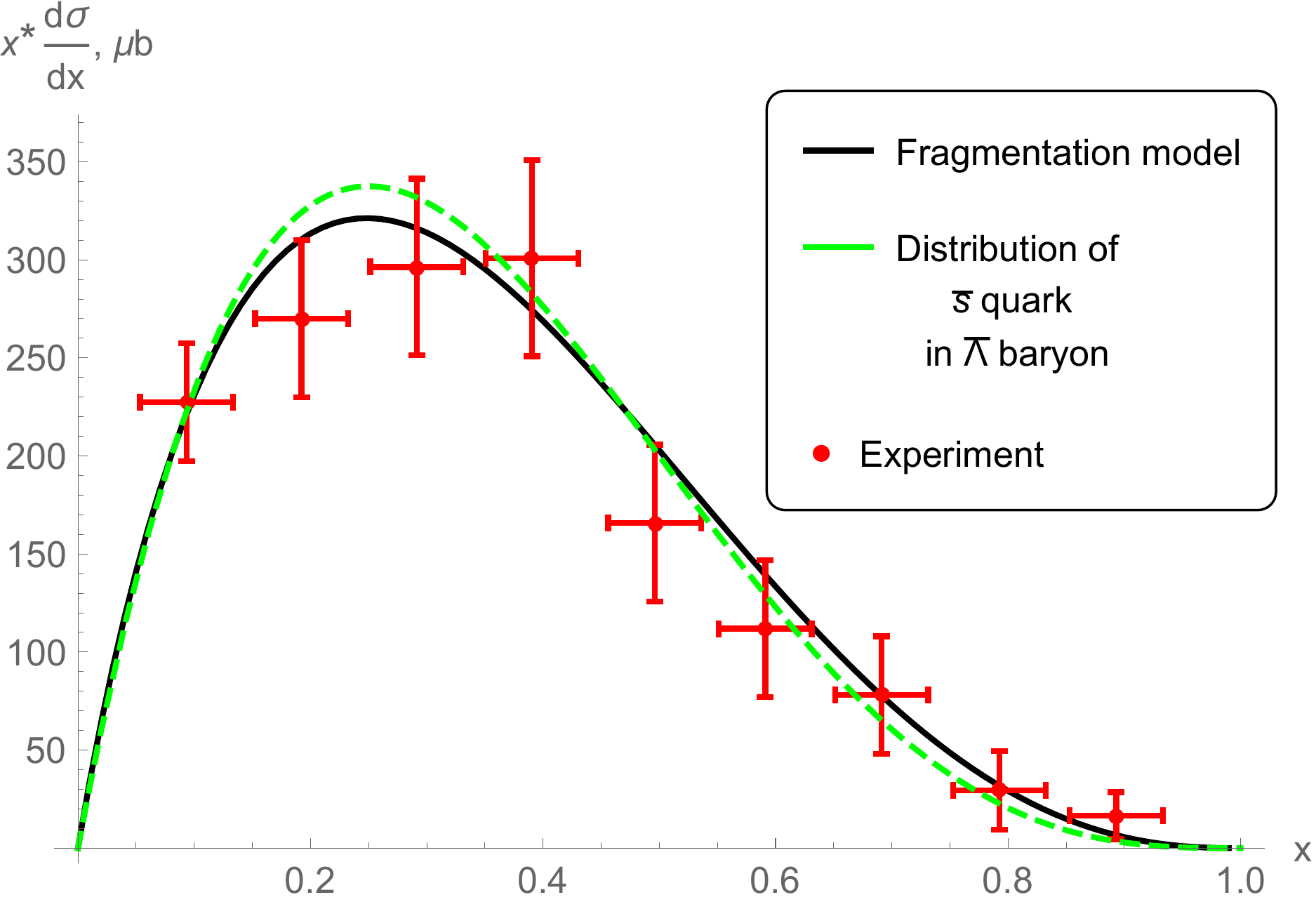}			
			\caption[]%
			{{\small $\bar{\Lambda}$}}    
		\end{subfigure}
		\hfill
		\begin{subfigure}[b]{0.475\textwidth}  
			\centering 
			\includegraphics[width=\textwidth]{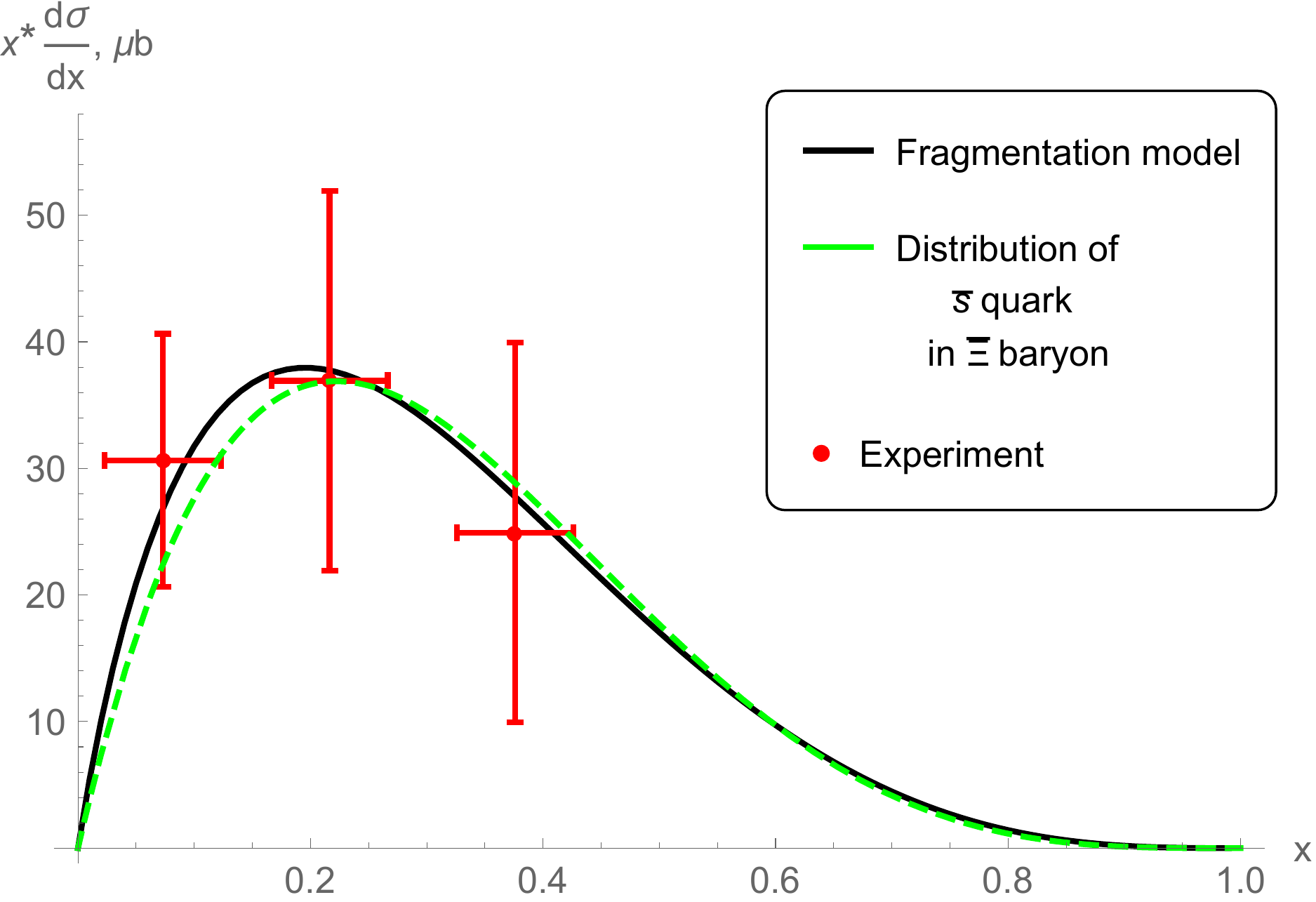}
			\caption[]%
			{{\small $\bar{\Xi}$}}    
		\end{subfigure}
		\caption[]
		{\small Differential cross section $x^*\frac{d\sigma}{dx}$ of $\bar{\Lambda}$ (a) and $\bar{\Xi}$ (b) baryons in the fragmentation model with $\gamma = 0.5 $ are shown in black, distribution of the valence quark in the baryon (green dotted line) and experimental data at 32 GeV, shown in red.} 
		\label{fig:frag_bar}
	\end{figure}

As in the case of a meson, when a baryon is produced, gluons do not have time to appear inside it, but taking into account the topology of the baryon, we assume that $\gamma_m$ does not vanish, but $\gamma_b=1$ and
	\begin{align}
		\phi^{q \to B} (z)= \frac{\Gamma(\gamma_b + 3 - \alpha_v - \beta_1 - \beta_2)}{\Gamma(1-\alpha_v)\Gamma(\gamma_b+2-\beta_1-\beta_2)}z^{-\alpha_v} * (1-z)^{2-\beta_1 -\beta_2}
	\end{align}
	Then in the case of fragmentation into a meson
	\begin{align}
		\label{eq:frag_mes_init_2}
		\frac{d\sigma}{dx} (x) = C_M \int_x^1 dy \ y^{-\alpha_v} * (1-y)^{\gamma_{m}-\alpha_2}\big( \frac{x}{y} \big)^{-\alpha_v}(1-\frac{x}{y})^{-\beta}, 
	\end{align}
	where
	\begin{align*}
		C_M = A^f \frac{\Gamma(2+\gamma_m-\alpha_v-\alpha_2)}{\Gamma(1-\alpha_v)\Gamma(1+\gamma_m-\alpha_2)} \frac{\Gamma(2+\gamma_m-\alpha_v-\beta)}{\Gamma(1-\alpha_v) \Gamma(1+\gamma_m-\beta)} 
	\end{align*}
	After the simple mathematical transformations (\ref{eq:frag_mes_init_2}) is converted to the form
	\begin{align}
		\frac{d\sigma}{dx} (x)  = C_M x^{-\alpha_v} \int_x^1 dy \ y^{\beta} * (1-y)^{\gamma_{m}-\alpha_2} (y-x)^{-\beta}.
	\end{align}
	For fragmentation into a baryon
	\begin{align}
	\label{eq:frag_bar_init_2}
		\frac{d\sigma}{dx} (x) = C_B \int_x^1 dy \ y^{-\alpha_v} * (1-y)^{\gamma_{m}-\alpha_2}\big( \frac{x}{y} \big)^{-\alpha_v}(1-\frac{x}{y})^{2-\beta_1-\beta_2}, 
	\end{align}
	where
	\begin{align*}
		C_B = A^f \frac{\Gamma(2+\gamma_m-\alpha_v-\alpha_2)}{\Gamma(1-\alpha_v)\Gamma(1+\gamma_m-\alpha_2)}  \frac{\Gamma(\gamma_b + 3 - \alpha_v - \beta_1 - \beta_2)}{\Gamma(1-\alpha_v)\Gamma(\gamma_b+2-\beta_1-\beta_2)}.
	\end{align*}
	After simplifications (\ref{eq:frag_bar_init_2}) becomes
	\begin{align}
		\frac{d\sigma}{dx} (x)  = C_B x^{-\alpha_v} \int_x^1 dy \ y^{\beta_1+\beta_2-2} * (1-y)^{\gamma_{m}-\alpha_2} (y-x)^{2-\beta_1-\beta_2}. 
	\end{align}
In Fig.~\ref{fig:frag_mes_xdsdx} the differential cross sections for the production of mesons $K^{*0}$, $\phi$ 
	 and their comparison with experiment at   energy $K^{+}p$ 250 GeV are presented.
The fragmentation model we are considering makes it possible for the first time to obtain theoretical predictions for inclusive baryon spectra.
In Fig.~\ref{fig:frag_bar} the differential cross sections for the production of $\bar{\Lambda}$, $\bar{\Xi}$ baryons in the fragmentation model for the case
$\gamma = 0.5$ and comparison with the distribution of the $\bar{s}$ valence quark in the corresponding baryon and experiment \cite{Azhinenko:1980cq}. 
\section{Tetraquark production}
\label{sect:ss}

Our analysis of particle production in a $K^+$-meson collision with a proton within the framework of our chosen models demonstrates a fairly satisfactory description of the spectra of ordinary hadrons, including the dominance of valence quarks and a weak dependence on the initial energy (scaling) in the fragmentation region (in our case, $ K$-meson).
	
\begin{figure}[h]
		\centering
		\begin{subfigure}[b]{0.475\textwidth}
			\centering
			\includegraphics[width=\textwidth]{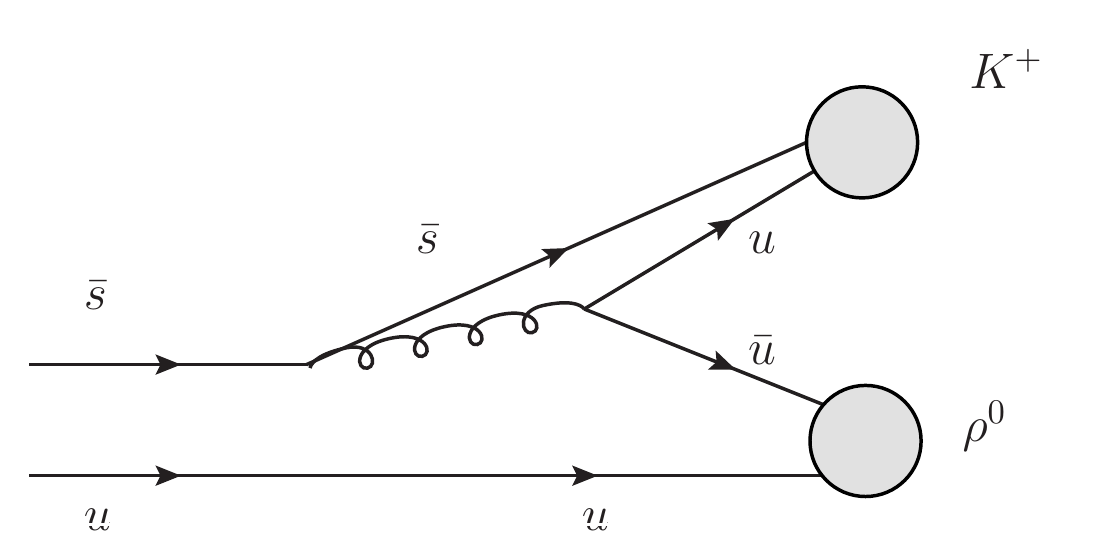}
			\caption[]%
			{{\small Production of a pair of mesons}}    
			\label{diag:pair_mes}
		\end{subfigure}
		\hfill
		\begin{subfigure}[b]{0.475\textwidth}  
			\centering 
			\includegraphics[width=\textwidth]{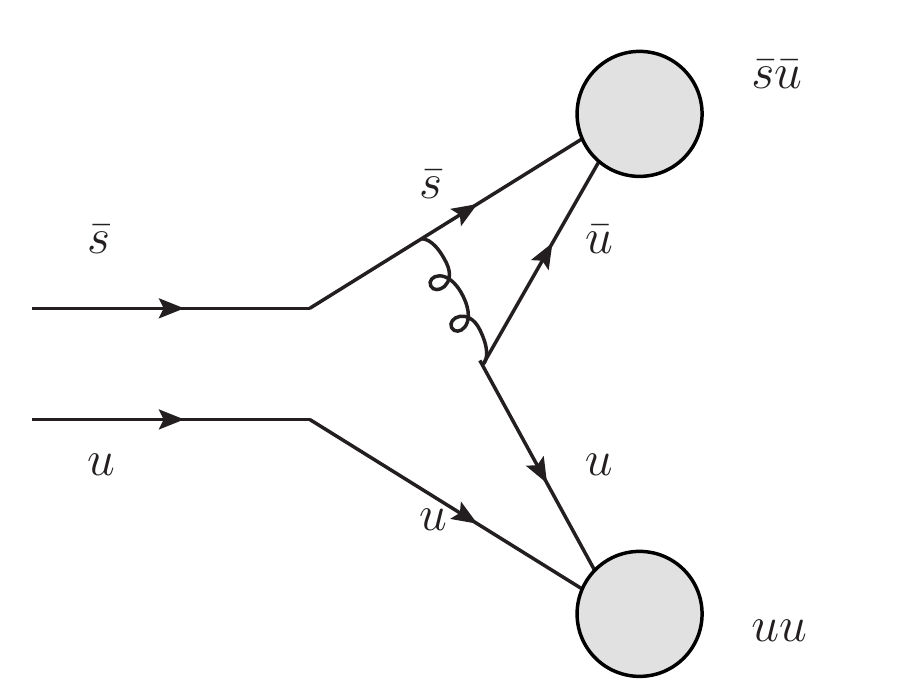}
			\caption[]%
			{{\small Production of a pair of diquarks}}
			\label{diag:pair_diquarks}
		\end{subfigure}
		\caption[]
		{\small Diagrams of the production of a pair of mesons (a) and a pair of diquarks (b)} 
		\label{diag:pair_mes_pair_diq}
\end{figure}

	Let us now turn to a discussion of the problem, which we indicated earlier, of elucidating the mechanism for the production of exotic states: a diquark (virtual) and a tetraquark. Let us consider the simplest cases of the formation of a hadron and a diquark. In Fig.~\ref{diag:pair_mes} the process of hadron production in the $K$-meson beam is presented, an example of the production of $K^+$- and $\rho^0$-mesons. We note that the formation of a whole spectrum of hadrons in each half of this diagram is possible, namely, states with different quantum numbers can be formed. For example, $K^{*+}$, $K^{**+}$, at the top of the diagram, and $\pi^0$, $f^0$, at the bottom.

\begin{figure}[h]
		\centering
		\includegraphics[width=0.5\linewidth]{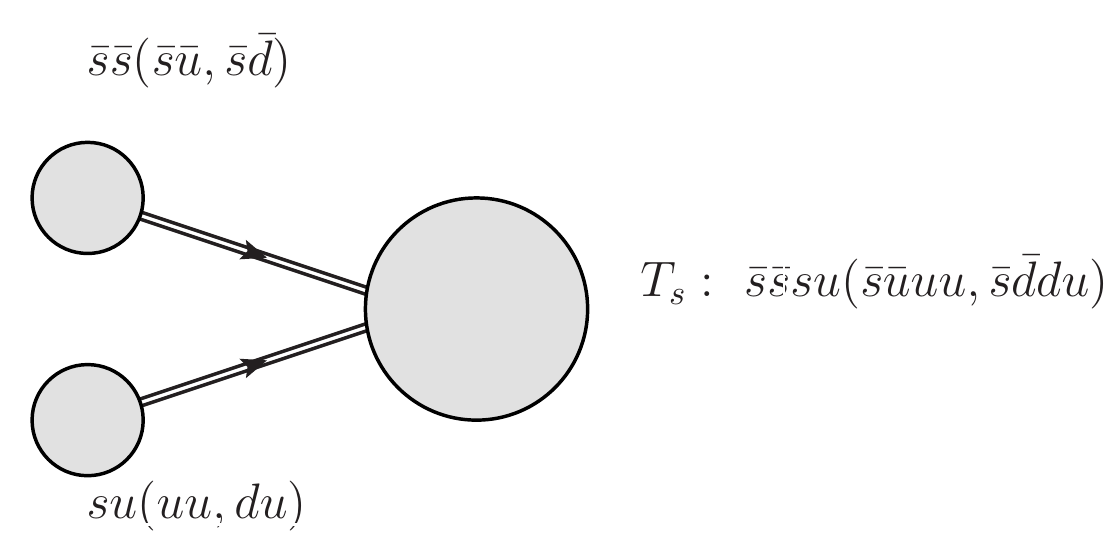}
		\caption{Diagram of the production of a doubly strange tetraquark}
		\label{diag:tetraquark_birth}
\end{figure}

The mechanism under consideration (Fig.~\ref{diag:pair_mes}) assumes decay into real physical states. Let us now consider the same picture, but with rearranged newly produced quarks (Fig.~\ref{diag:pair_diquarks}).
It can be seen that the initial colorless state has turned into two color states, each of which can form a bound state (attraction takes place in the triplet representation), and these bound states, two diquarks, can form a tetraquark, the mass spectrum of which can be obtained within the framework of the potential model or QCD sum rules. A diagram of the possible formation of a tetraquark with two strange quarks is shown in Fig.~\ref{diag:tetraquark_birth}.
	
	To calculate the mass of a tetraquark, one often confines oneself to the picture of the interaction of a diquark-anti-diquark
	pair, and the finite sizes of diquarks are taken into account by introducing a form factor. 
	There is also a more general approach based on the use of QCD sum rules. 
	The predicted mass spectrum for the lowest states is given in the Appendix~\ref{app:const}, 
	Tables \ref{tab:decays_tetra_suuu}, 
	\ref{tab:decays_tetra_sddu}, \ref{tab:decays_tetra_sssu}, and the actually observed (presumably exotic) 
	ones are given in same application in the Table~\ref{tab:exot_part} .
	
\begin{figure}[h]
		\centering
		\begin{subfigure}[b]{0.475\textwidth}
			\centering
			\includegraphics[width=\textwidth]{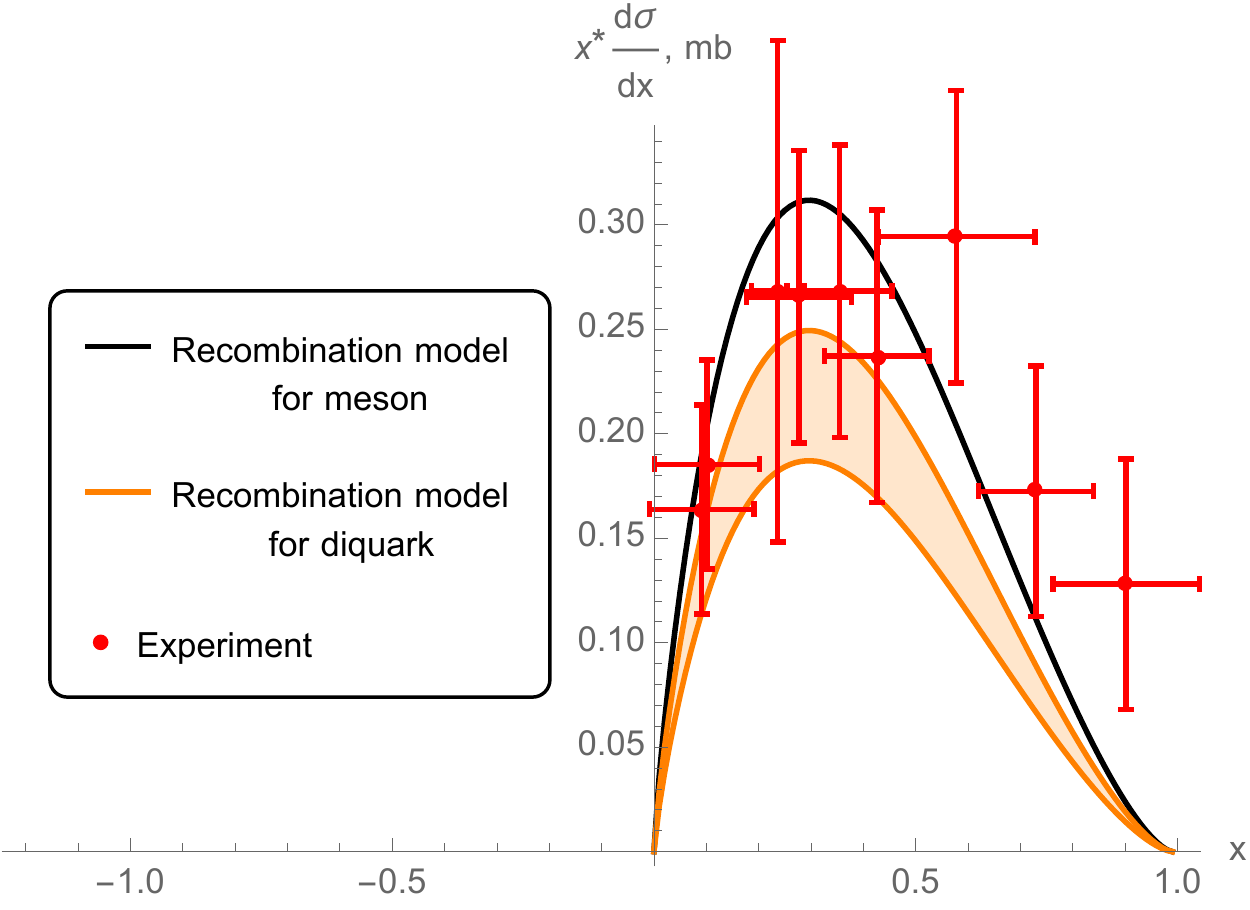}
			\caption[]%
			{{\small Recombination model. $\phi$ and $\bar{s}\bar{s}$}}    
		\end{subfigure}
		\hfill
		\begin{subfigure}[b]{0.475\textwidth}  
			\centering 
			\includegraphics[width=\textwidth]{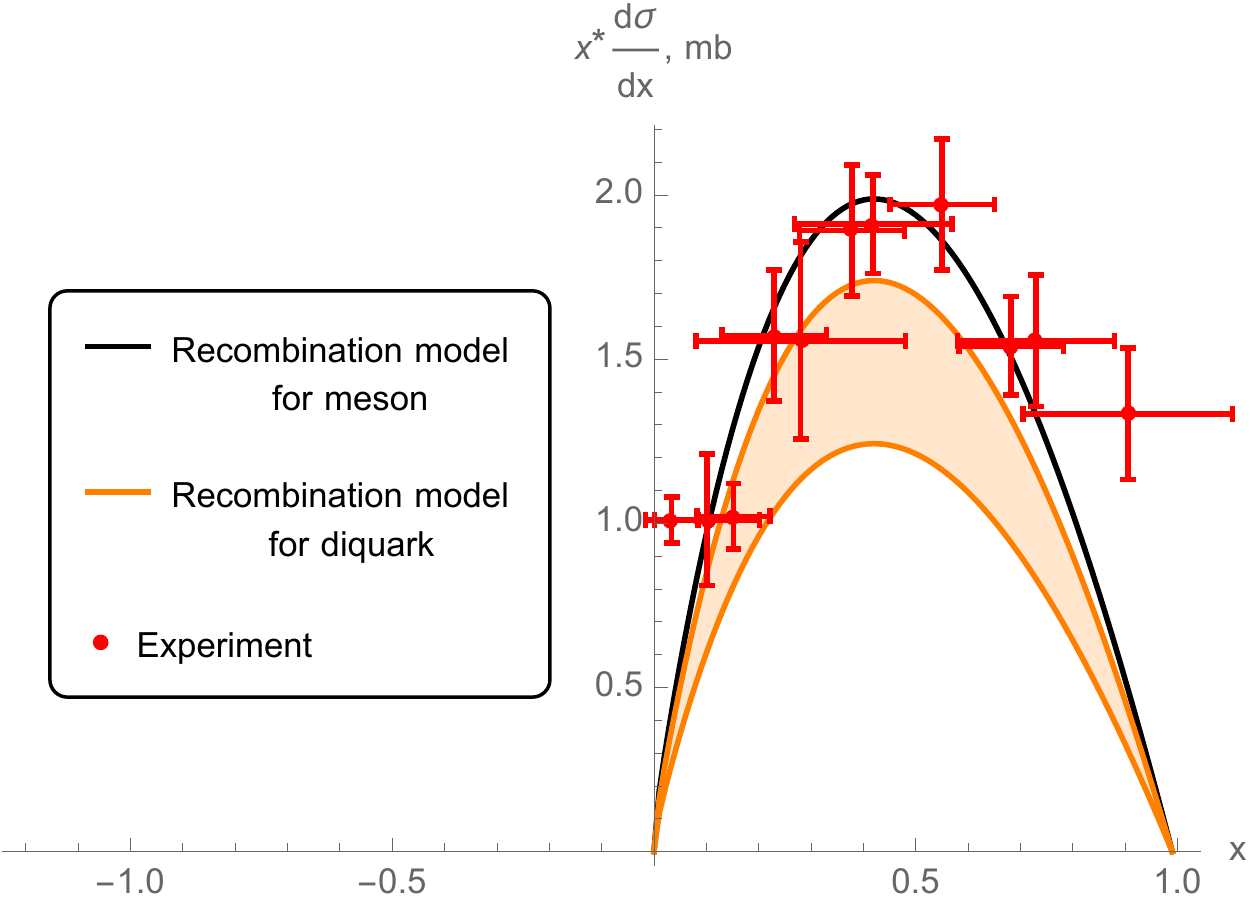}
			\caption[]%
			{{\small Recombination model. $K^{*0}$ and $\bar{s}\bar{d}$}}
		\end{subfigure}
		
		\vskip\baselineskip
		\begin{subfigure}[b]{0.475\textwidth}
			\centering
			\includegraphics[width=\textwidth]{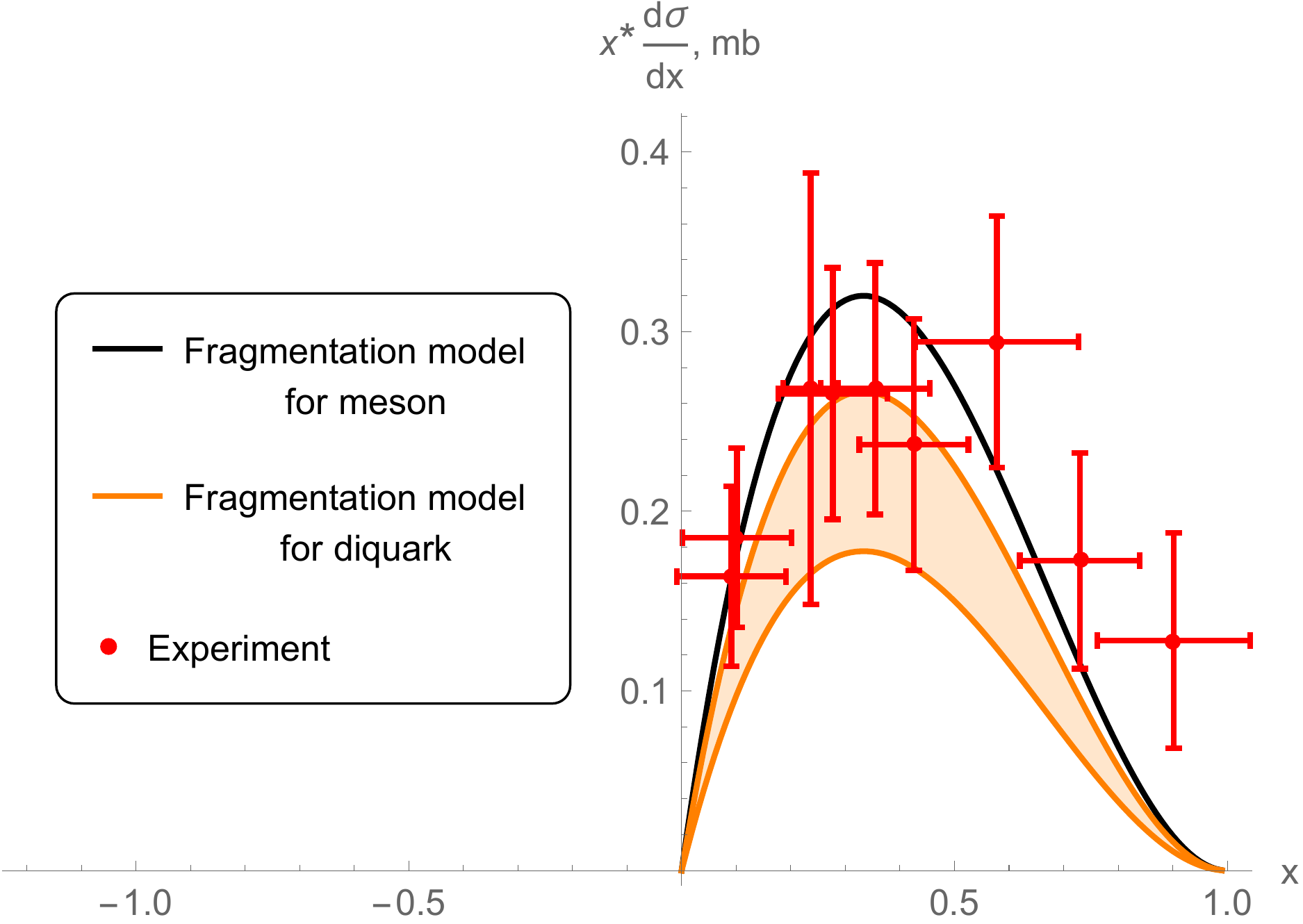}
			\caption[]%
			{{\small Fragmentation model. $\phi$ and $\bar{s}\bar{s}$}}    
		\end{subfigure}
		\hfill
		\begin{subfigure}[b]{0.475\textwidth}  
			\centering 
			\includegraphics[width=\textwidth]{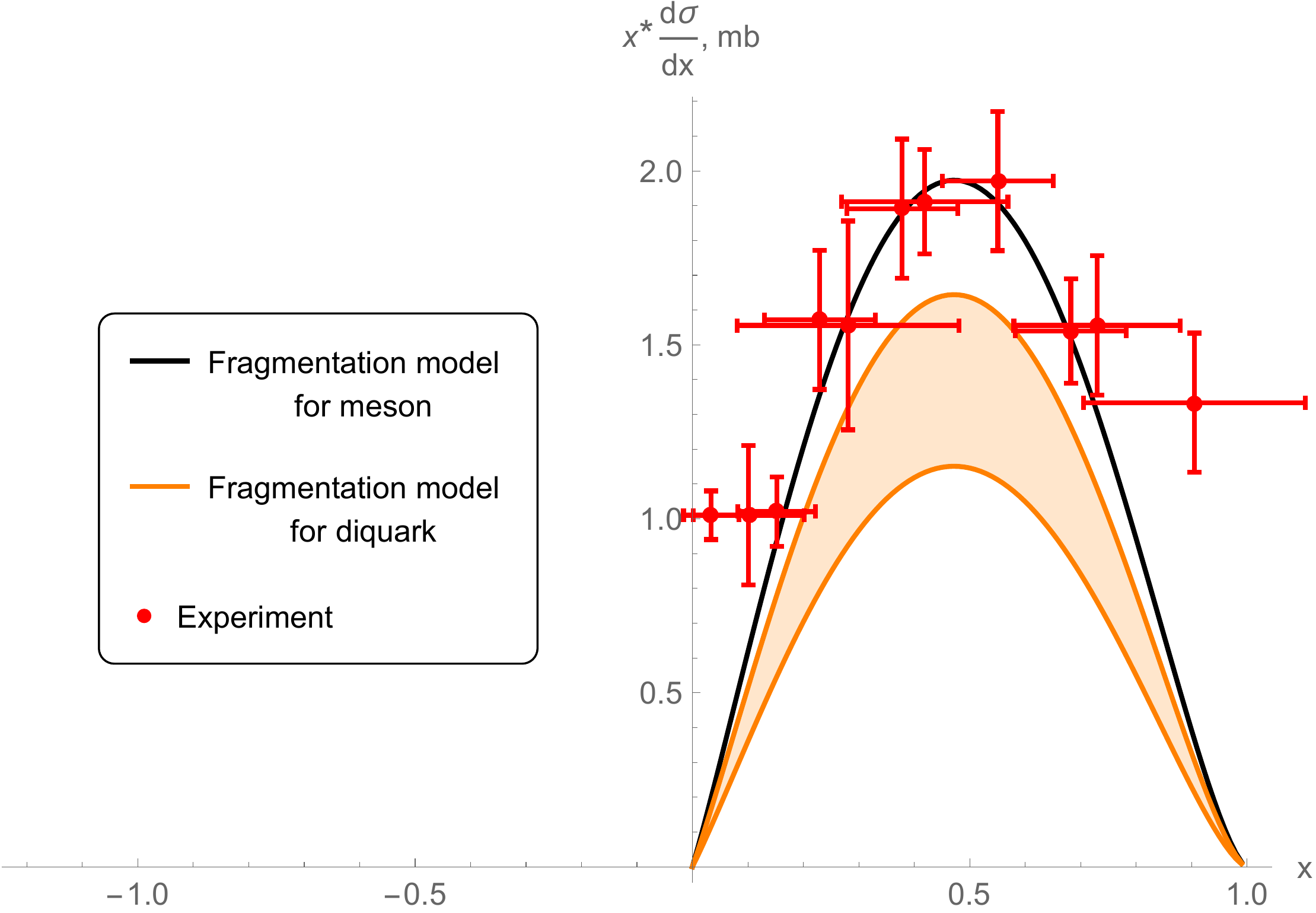}
			\caption[]%
			{{\small Fragmentation model. $K^{*0}$ and $\bar{s}\bar{d}$}}
		\end{subfigure}
		\caption[]
		{\small Comparison of differential cross sections for mesons (black lines) and diquarks (orange lines) in the recombination (a,b) and fragmentation (c,d) models} 
		\label{fig:diquark_meson_sravn}
	\end{figure}

	The inclusive production spectra of $\bar{\Lambda}$ and $\bar{\Xi}$ presented earlier (Fig. \ref{fig:frag_bar} ) are satisfactorily described in the fragmentation model. An important question that arises when observing these antibaryons is the law of conservation of baryon number, which in each case implies the birth of a baryon partner. In our opinion, the only reasonable mechanism for the production of a baryon pair is the decay of a tetraquark into a baryon-antibaryon pair.
	The formation of tetraquarks requires the preliminary production of diquarks, which are formed in the $K^+$ beam along with mesons. The diquark production diagram is shown above in Fig. \ref{diag:pair_diquarks}.
	The production of diquarks is very similar to the production of mesons.
Our estimates show that the production spectra of $\bar{s}\bar{d}$ and $\bar{s}\bar{s}$ vector diquarks should repeat the spectra of $K^{*0}$- and $\phi $-meson, respectively, with a comparable cross section.
In Fig.~\ref{fig:diquark_meson_sravn} the differential cross sections for diquark and meson production 
predicted by the recombination model and the fragmentation model for $\bar{s}\bar{s}$, $\bar{s}\bar{d}$ are compared diquarks
and $\phi$-, $K^{*0}$-mesons, respectively.	
	Based on the distribution of the cross section of diquarks, we can conclude that, along with mesons, a sufficient number of diquarks are produced in the $K^+$ beam to speak of the production of a pair of diquarks that have attraction and form a tetraquark that can be observed experimentally.
	
\begin{figure}[h]
		\centering
		\begin{subfigure}[b]{0.475\textwidth}
			\centering
			\includegraphics[width=\textwidth]{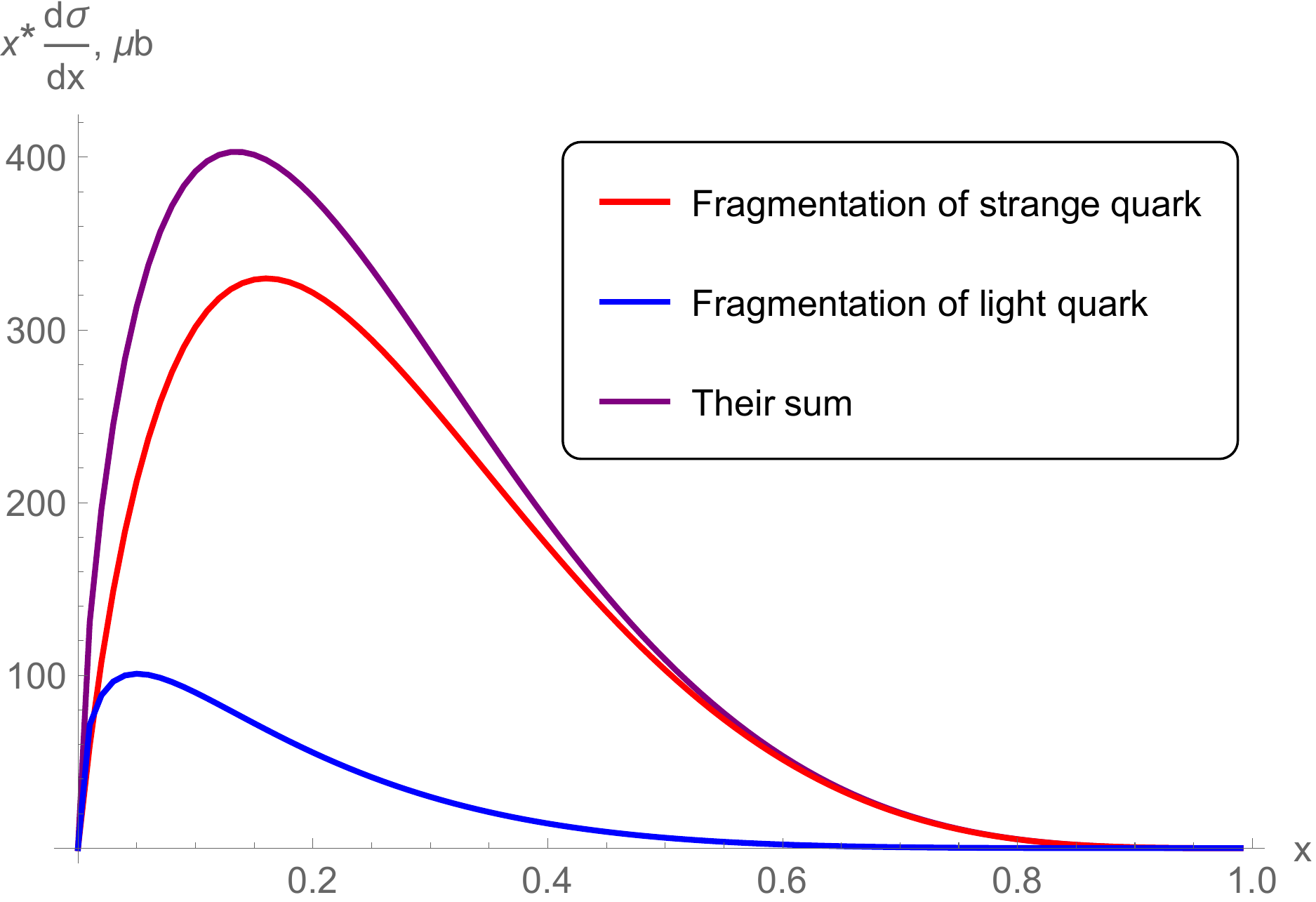}
			\caption[]%
			{{\small $\bar{s}\bar{u}uu$ ($\bar{s}\bar{d}du$) }}    
		\end{subfigure}
		\hfill
		\begin{subfigure}[b]{0.475\textwidth}  
			\centering 
			\includegraphics[width=\textwidth]{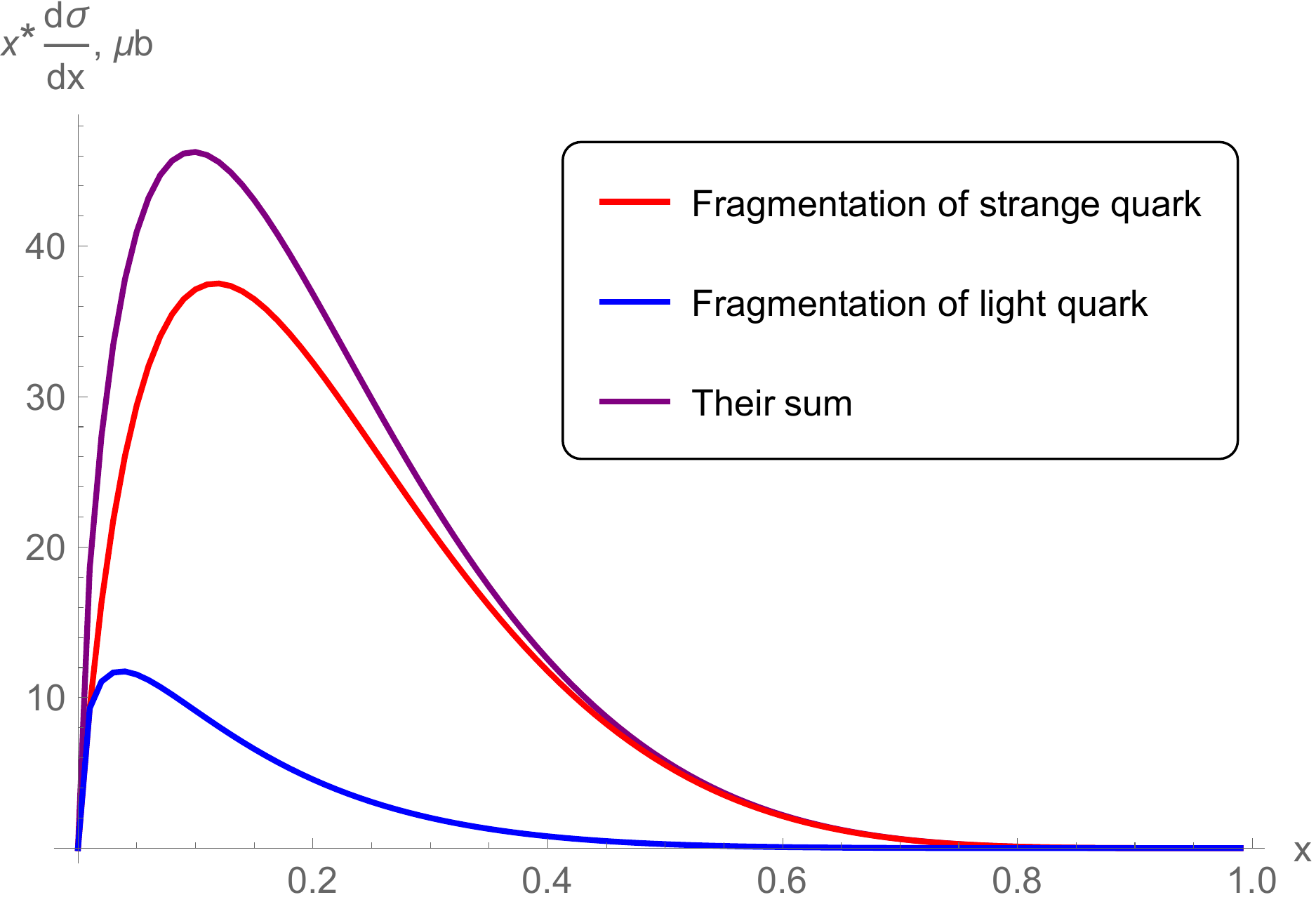}
			\caption[]%
			{{\small $\bar{s}\bar{s}su$}}
		\end{subfigure}
		\caption[]
		{\small Tetraquark production diagrams $\bar{s}\bar{u}uu$ (a), $\bar{s}\bar{d}du$ (b) and $\bar{s}\bar{s}su$ (c) in the fragmentation model. The red curves are the fragmentation of the strange quark, the blue curves are the fragmentation of the light quark, the purple curves are their sum.} 
		\label{pic:tetraquark_fragmentation}
	\end{figure}

\begin{figure}[h]
		\centering
		\begin{subfigure}[b]{0.475\textwidth}
			\centering
			\includegraphics[width=\textwidth]{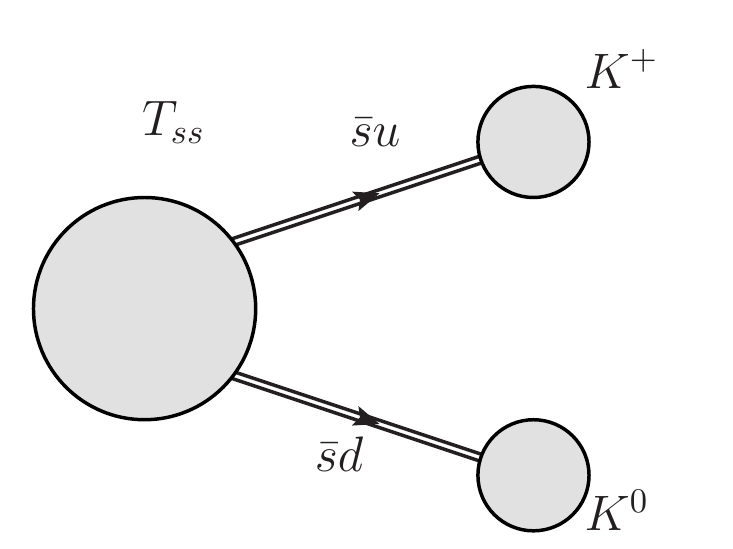}
			\caption[]%
			{{\small Decays into a pair of mesons}}    
		\end{subfigure}
		\hfill
		\begin{subfigure}[b]{0.475\textwidth}  
			\centering 
			\includegraphics[width=\textwidth]{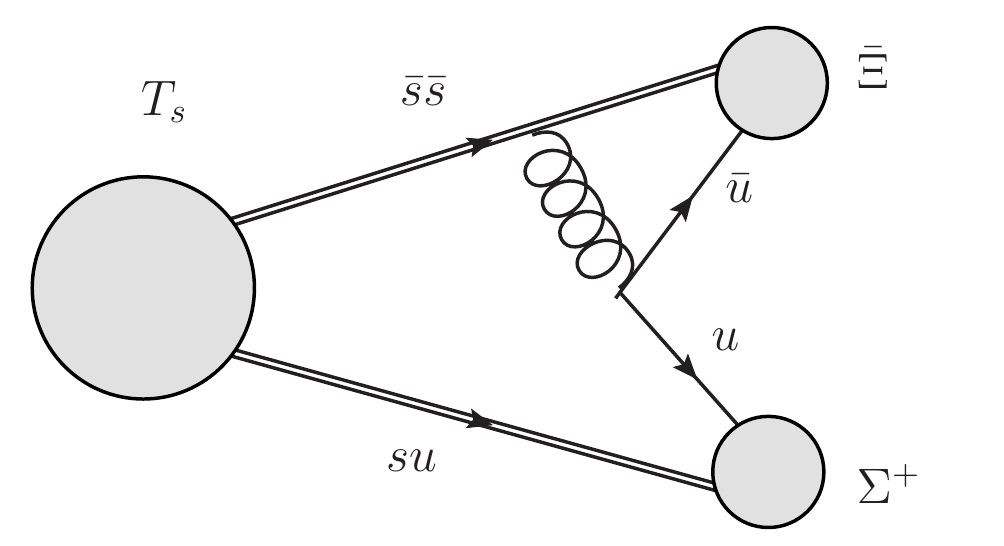}
			\caption[]%
			{{\small Decays into a pair of baryons}}
		\end{subfigure}
		
		\caption[]
		{\small Decays of tetraquarks into a pair of mesons (a) and a pair of baryons (b) } 
		\label{diag:tetraquark_decays}
	\end{figure}

	In Fig.~\ref{pic:tetraquark_fragmentation} presents our theoretical predictions for the differential cross section for the production of $\bar{s}\bar{u}uu$, $\bar{s}\bar{d}du$ and $\bar{s}\bar{s}su$ in the fragmentation model.
In the case of the birth of a tetraquark, it is quite obvious that it can decay into a pair of mesons. Particularly interesting is the fact that a decay channel into a baryon-antibaryon pair is also possible.
Diagrams of these processes are presented in Fig.~\ref{diag:tetraquark_decays}.
%
\section{How to look for tetraquarks?}
\label{sect:find}

So far, it has been possible to experimentally detect tetraquarks with sufficient certainty only in the heavy quark sector. In the light quark sector, the situation is much poorer, but there are still a number of exotic resonances, some of which can claim to be tetraquark states. A significant contribution to the search for these resonances was made by the BESIII experiment, during which, for example, the resonant states $X(2120)$, $X(2370)$ were found in $J/\psi \to \gamma \pi^+ \pi ^- \eta^{\prime}$ and $X(2500)$ in the $\phi \phi$ mass spectrum in the $J/\psi \to \gamma \phi \phi$ decay. Detailed information about exotic resonances is presented in the \ref{tab:exot_part} table. Note that the range of particle masses in this table lies in the range 2100 --- 2500 MeV. As for the widths of the resonances, they are present as relatively narrow (the width of $X(2120)$ is $\approx$ 83~MeV),
and quite wide (for $X(2500) \approx$ 230~MeV).
It is difficult to conclude whether these resonances are states of two or four quarks. Nevertheless, in the examples $X(2120)$, $X(2370)$, and $X(2500)$ presented by us, we consider the four-quak nature of the resonances to be the most probable. First,
based on the works \cite{Lu:2019ira, Liu:2020lpw, Su:2020reg, Li:2020xzs}
(which we will talk about later), we can expect that decays into two or more resonances dominate in the decays of states of tetraquarks with lower spins. Secondly, it is difficult to imagine a two-quark state decaying into a system $\pi^- \eta^{\prime}$ or $\phi \phi$.   

One of the main goals of our work is to evaluate the possibility of an experimental search for tetraquarks in a $K^+$ meson beam. The presence of such a beam makes it possible to deal with a more probable process of the production of strange particles due to the presence of a ready-made strange quark. We believe that due to this, the production of tetraquarks $\bar{s}\bar{u}uu$, $\bar{s}\bar{d}du$ and $\bar{s }\bar{s}su$ in sufficient quantity for experimental detection.

As for the estimation of the mass spectrum for states of light tetraquarks with different quantum numbers, there are a large number of predictions \cite{Lu:2019ira, Liu:2020lpw, Su:2020reg, Li:2020xzs}, some of which we use for our estimates.
In particular, \cite{Liu:2020lpw} predicts the mass spectrum of a completely strange tetraquark within the nonrelativistic potential quark model (see Fig. 2 from \cite{Liu:2020lpw}). According to these predictions, the mass $1S$ of the state of a completely strange tetraquark
lies in the range 2218 -- 2440 MeV, $1P$ states --- in the range 2445 -- 2984 MeV, $2S$ states --- in the range 2798 -- 3155 MeV.

Among the tetraquarks $\bar{s}\bar{u}uu$, $\bar{s}\bar{d}du$ and $\bar{s}\bar{s}su$ of interest to us, there are still no \-pre\-diction
for the tetraquark $\bar{s}\bar{s}su$. As regards tetraquarks containing one strange quark and three light quarks, it is assumed in \cite{Ebert:2008id} that the mass of such tetraquarks is in the range of $700 - 2000$~MeV. Note that, most likely, the lower mass limit in this work is underestimated, because predictions do not take into account that diquarks in a tetraquark are not point-like, but are "smeared" in space, which increases their mass.

For the tetraquark $\bar{s}\bar{s}su$, we have the opportunity to roughly estimate the range of masses of various states using the estimates for the completely strange tetraquark and the smaller constituent mass of light quarks $u$, $d$ in comparison with the strange quark in hadron.

As we said earlier in this paper, pairs of mesons and pairs of diquarks are produced in the $K$-meson beam (Fig. \ref{diag:pair_mes_pair_diq}), and the latter can then form a tetraquark state. We expect this state to decay into a pair of meson resonances with a high probability. In some cases, such decays make it possible to determine the tetraquark nature of the resonances. As stated earlier, we expect tetraquarks to decay into multiple resonant states. The simplest way for the decay of the tetraquarks we are considering is precisely the decay into two mesons. An example of such a decay is the $\bar{s}\bar{s}su \to \phi K^{*+}$ decay. In this case, as before for the $\pi^- \eta^{\prime}$ and $\phi \phi$ systems, it is difficult to imagine that the $\phi K^{*+}$ system was produced from a two-quark resonance.

The decay of tetraquarks into a baryon-antibaryon system is very interesting. Such a decay is possible if one of the quarks in a tetraquark emits a gluon that forms a quark-antiquark pair (see Fig. \ref{diag:tetraquark_decays}b).
Note that there are a number of papers \cite{Haidenbauer:2023zcu, BESIII:2019tgo, Achasov:2014ncd, BESIII:2021fqx} where the production of baryon-antibaryon pairs is considered. For example, in the reactions $e^+ e^- \to \phi \Lambda \bar{\Lambda}$ and $e^+ e^- \to \eta \Lambda \bar{\Lambda}$ \cite{Haidenbauer:2023zcu}. We assume that the pair $\Lambda \bar{\Lambda}$ is produced from the tetraquark $\bar{s}\bar{u}su$ or $\bar{s}\bar{d}sd$.

Relying on the pair production of baryons in the experiment, we assume that the baryons $\bar{\Lambda}$, $\bar{\Xi}$ in the \cite{Azhinenko:1980cq} experiment were also produced precisely from decays of tetraquarks. In this case, the baryon $\bar{\Lambda}$ was produced from the decay of $\bar{s}\bar{u}uu \to \bar{\Lambda} X$ or $\bar{s}\bar{d} du \to \bar{\Lambda} X$, where $X$ is a proton or a delta isobar. Then the cross section of the process $\bar{s}\bar{u}uu \to \bar{\Lambda} p$ can be estimated from above by the production cross section $\bar{\Lambda}$. The baryon $\bar{\Xi}$, in turn, is produced from the decays of the tetraquark $\bar{s}\bar{s}su$.
This fact allows us to estimate the cross section for the production of tetraquarks from below by the cross section for the production of antibaryons. Then for the tetraquark $\bar{s}\bar{u}uu$ ($\bar{s}\bar{d}du$) the cross section is $\sigma \geq 420$~$\mu$b, and for the tetraquark $\bar{s }\bar{s}su$ --- $\sigma \geq 36$~$\mu$b.
The expected most probable decay modes for the $\bar{s}\bar{u}uu$, $\bar{s}\bar{d}du$, $\bar{s}\bar{s}su$ tetraquarks are presented in the Tables \ref{tab:decays_tetra_suuu}, \ref{tab:decays_tetra_sddu}, \ref{tab:decays_tetra_sssu} .
\section{Conclusion} 
  
As already noted in the Introduction, studies related to light multiquarks open up an interesting new perspective in the region of nonperturbative QCD.
Research in this area is associated with significant difficulties both in relation to the mechanisms of production of light multiquarks and their spectral characteristics (mass and width).
It is clear that in-depth research along this path will require the development and use of new specific methods.
As a first step, we chose well-established nonperturbative models that allowed us to obtain a number of specific results that could attract the attention of experimenters.
In particular, it is proposed that the search for light tetraquarks could be carried out using the $K$ beams of the IHEP accelerator. In this connection, we have given estimates of the inclusive cross sections for a number of processes at a $K$ beam energy of 32~GeV.
\section*{Acknowledgments}

The authors are grateful to P. V. Chliapnikov, S. P. Denisov, V. F. Obraztsov, A. M. Zaitsev
and participants of the seminar of the Division of experimental physics for their attention and useful discussions. 
%
\section{Appendix} 
\subsection{Distribution of quarks in the $K^+$-meson and in the proton, the beam  $K^+ p$,  $\lambda_s = 0.3$}

\label{app:part_raspr}
	\begin{align}
		&V_{u}^{p} = 1.79x^{-1/2}(1+2.3x)(1-x)^3 \nonumber \\
		&V_{d}^{p} = 1.1x^{-1/2}(1-x)^{3.1} \nonumber \\
		&S_{u, \bar{u}, d, \bar{d}}^{p} = 0.3x^{-1}(1-x)^7 \nonumber \\
		&S_{s,\bar{s}}^{p} = \lambda_s 0.3x^{-1}(1-x)^7 \nonumber\\
		&V_{s,\bar{s}}^{K} = 2(1-x) \nonumber \nonumber \\
		&S_{u, \bar{u}, d, \bar{d}}^{K} = 0.3x^{-1}(1-x)^5 \nonumber \\
		&S_{s,\bar{s}}^{K} = \lambda_s 0.3x^{-1}(1-x)^5 \nonumber
	\end{align}

\label{app:const}
	\begin{table}[h]
		\begin{center}
			\begin{tabular}{|c|c|c|}
				\hline
				\textbf{Quantum numbers} & \textbf{Decay into a pair of mesons} & \textbf{Decay into a pair of baryons} \\
				\hline
				$0^{++}$ &  $K^+ \pi^0 K^{*+} \rho^0, K^{*0} \rho^+$ & $\bar{\Lambda} p, \bar{\Sigma}^{*-} \Delta^{++}$ \\
				\hline
				$1^{+-}$ &  $K^{+} \rho^0, K^{*+} \pi^0$ & $\bar{\Lambda} \Delta^{+}, \bar{\Lambda}^* p$\\
				\hline
				$2^{++}$ &  $K^+ \pi^0, K^{*+} \rho^0,  K^+ \rho^0$ & $\bar{\Lambda} p$ \\
				\hline
			\end{tabular}
			\caption{Predictions for Tetraquark Decays $uu\bar{u}\bar{s}$ in the 1S State}
			\label{tab:decays_tetra_suuu}
		\end{center}
	\end{table}
	\begin{table}[h]
		\begin{center}
			\begin{tabular}{|c|c|c|}
				\hline
				\textbf{Quantum numbers} & \textbf{Decay into a pair of mesons} & \textbf{Decay into a pair of baryons} \\
				\hline
				$0^{++}$ &  $K^0 \pi^+, K^{*0} \rho^+, K^{*0} \rho^+$ & $\bar{\Lambda} p, \bar{\Sigma}^{*-} \Delta^{++}$ \\
				\hline
				$1^{+-}$ &  $K^{0} \rho^+, K^{*0} \pi^+$ & $\bar{\Lambda} \Delta^{+}, \bar{\Lambda}^* p$\\
				\hline
				$2^{++}$ &  $K^0 \pi^+, K^{*0} \rho^+,  K^0 \rho^+$ & $\bar{\Lambda} p$ \\
				\hline
			\end{tabular}
			\caption{Predictions for Tetraquark Decays $ud\bar{d}\bar{s}$ in the 1S State}
			\label{tab:decays_tetra_sddu}
		\end{center}
	\end{table}
	\begin{table}[h]
		\begin{center}
			\begin{tabular}{|c|c|c|}
				\hline
				\textbf{Quantum numbers} & \textbf{Decay into a pair of mesons} & \textbf{Decay into a pair of baryons} \\
				\hline
				$0^{++}$ &  $\phi K^{*+}$ & $\bar{\Xi} \Sigma^+ $ \\
				\hline
				$1^{+-}$ &  $\eta K^{*+}, \eta^{\prime}K^{*+}$ & $\bar{\Xi} \Sigma^{*+}, \bar{\Xi}^* \Sigma^+ $\\
				\hline
				$2^{++}$ &  $\phi K^{*+}, \eta K^{+}, \eta^{\prime}K^{+}$ & $\bar{\Xi} \Sigma^+$ \\
				\hline
			\end{tabular}
			\caption{Predictions for Tetraquark Decays $us\bar{s}\bar{s}$ in the $1S$ State}
			\label{tab:decays_tetra_sssu}
		\end{center}
	\end{table}
	\begin{table}[h]
		
		\begin{center}
			\begin{tabular}{|c|c|c|c|c|}
				\hline
				\textbf{Hadron} & \textbf{$J^{PC}$} & \textbf{Experiment, year} & \textbf{Mass, MeV} & \textbf{Width, MeV}  \\
				\hline
				Y(2175) & $1^{--}$ & Babar, 2006, $e^+e^- \to \phi f_0(980)$  & $2175 \pm 10 \pm 15$ &  $58 \pm 16 \pm 20$  \\
				\hline
				Y(2175) &  &BESII, 2008  & $2186 \pm 10 \pm 6$ & $65 \pm 23 \pm 17$   \\
				\hline
				X(2400) &  &Belle and BaBar, 2010 & $2436 \pm 26$ & $121 \pm 35$    \\
				\hline
				X(2240)&  & BESIII, 2019 $e^+e^- \to K^+ K^- $    & $2239 \pm 7.1 \pm 11.3$ & $139 \pm 12.3 \pm 20.6$  \\
				\hline
				X(2300)& & BESIII, 2021 $e^+e^- \to \phi \pi^+ \pi^-$ & $2298^{+60}_{-44} \pm 6$ & $219^{+117}_{-112}\pm 6$ \\
				\hline
				X(2360)& & BESIII, 2022 $e^+e^- \to \Lambda \bar{\Lambda} \eta $& $2356 \pm 7 \pm 17$ & $304 \pm 28 \pm 54$ \\
				\hline
				X(2120)&  & BESIII, 2010 $J/\psi \to \gamma \pi^+ \pi^- \eta^{\prime}$ & $2122 \pm 6.7^{+4.7}_{-2.7}$ & $83 \pm 16 ^{+31}_{-11}$ \\
				\hline
				X(2370) &  & BESIII, 2010 $J/\psi \to \gamma \pi^+ \pi^- \eta^{\prime}$ & $2376.3 \pm 8.7^{+3.2}_{-4.3}$ & $83 \pm 17^{+44}_{-6}$ \\
				\hline
				X(2500) & $0^-$ & BESIII, 2016 $J/\psi \to \gamma \phi \phi$ & $2470^{+15+101}_{-19-23}$ & $230^{+64+56}_{-35-33}$ \\
				\hline
				
			\end{tabular}
			
			\caption{Exotic resonances}
			\label{tab:exot_part}
		\end{center}
	\end{table}	
	
%
%
\newpage
\newpage
\bibliography{Tetra.bib}

%
\end{document}